\UseRawInputEncoding
\documentclass[
superscriptaddress,
bibnotes,
 amsmath,amssymb,
 aps,
pra,
twocolumn,
longbibliography
floatfix,
]{revtex4-1}
\usepackage{graphicx}
\usepackage{dcolumn}
\usepackage{xr}
\usepackage{bm} 
\usepackage{float}
\usepackage{hyperref} 
\hypersetup{
	colorlinks=true,
	linkcolor=blue,
	citecolor=blue,
	filecolor=magenta,
	urlcolor=blue         
}
\usepackage{color,soul}
\usepackage{lineno} 

\renewcommand{\selectlanguage}[1]{}

\begin{document}

\title{Frequency-entangled $W$ states and quantum frequency translation protocols via forward Brillouin interactions}

\author{Andrew J. Shepherd}
\email[]{as5262@nau.edu}
\address{Department of Applied Physics and Materials Science, Northern Arizona University, Flagstaff, AZ 86011, USA}
\address{Center for Materials Interfaces in Research and Applications, Flagstaff, AZ, USA}

\author{Ryan O. Behunin}
\email[]{ryan.behunin@nau.edu}
\address{Department of Applied Physics and Materials Science, Northern Arizona University, Flagstaff, AZ 86011, USA}
\address{Center for Materials Interfaces in Research and Applications, Flagstaff, AZ, USA}

\date{\today}

\begin{abstract}
Complex quantum states of light are not only central to advancing our understanding of quantum mechanics, but are also necessary for a variety of quantum protocols. High-dimensional, or multipartite, quantum states are of specific interest, as they can exhibit unique properties both fundamentally and in application. The synthesis of high-dimensional, entangled photonic states can take the form of various schemes, which result in varying forms of entanglement. Frequency-entanglement is specifically attractive due to compatibility with integrated systems and resistance to decoherence in fiber transportation; however, increasing the dimension of frequency-entangled states requires a system that offers quantum interactions between a large set of distinct frequencies. Here, we show how the phonon-photon interactions of forward Brillouin scattering, which offer access to a ladder of optical resonances permitted by a single mechanical mode, can be used for fast-synthesis of frequency-entangled, single-photon $W$ states. In our proposed system, simultaneous laser pulses of different frequencies dynamically evolve either an injected single photon or a heralded single phonon, generating $W$ states of selected dimension and output frequency. This method enables the synthesis of `perfect' $W$ states by adjusting the pulse amplitudes. In addition, we show how this system can be used for quantum frequency translation.
\end{abstract} 

\maketitle

\section{Introduction}
Entangled quantum states of light are a natural candidate for quantum protocols \cite{yuan_entangled_2010, flamini_photonic_2018} and tests of fundamental physics \cite{guhne_entanglement_2009} as they are fast, easily manipulable, and relatively robust against decoherence. As the depth of quantum information processing grows, many applications require multipartite entangled states of increased dimension and complexity \cite{walter_multipartite_2016,gimeno-segovia_three-photon_2015,cleve_substituting_1997,raussendorf_one-way_2001}. This necessitates practical protocols for fast, high-fidelity generation of multipartite states that are scalable to large dimensions, resistant to loss, and compatible with state of the art technologies, such as contemporary optical fibers and quantum memories. Bipartite photonic states, including maximally entangled Bell-states, have been synthesized and studied extensively \cite{flamini_photonic_2018}. Extending to higher-dimensional multipartite states is of specific interest as they can exhibit unique qualities \cite{walter_multipartite_2016}, such as new conflicts with local realism compared to their bipartite counterparts \cite{greenberger_going_1989}, and enhancement of quantum computation and communication efficiency \cite{gimeno-segovia_three-photon_2015,cleve_substituting_1997}. $W$ states are one of two paradigmatic examples of multipartite entangled states, in addition to GHZ-states. The maximally entangled, standard $W$ state is defined as 
\begin{align}
\label{eq:standard-W-state}
    |W_N\rangle = \frac{1}{\sqrt{N}}\bigl(|100...0\rangle + |010...0\rangle + ... + |00...01\rangle \bigr),
\end{align}
where $\{|0\rangle,|1\rangle\}$ is the orthonormal basis of $N$-physical or logical qubits, which could describe the ground/excited states of a two-level systems, the vacuum/single excitation of bosonic modes, or the horizontal/vertical polarization of photons, for example.

$W$ states are attractive for being robust against loss \cite{dur_three_2000}, as tracing out a subsystem preserves the maximum amount of entanglement compared to any other $N$-dimensional state. This has led to their investigation for a variety of quantum protocols, such as teleportation \cite{joo_quantum_2003}, quantum cryptography \cite{jian_quantum_2007}, quantum information processing \cite{singh_usefulness_2016}, analysis of nonlocality \cite{cabello_bells_2002}, and even increasing interferometer baseline lengths in telescope arrays \cite{gottesman_longer-baseline_2012}. Some protocols may require a different type of $W$ state, where the coefficients in each ket to differ from Eq. \eqref{eq:general_w}. One case is the `perfect' $W$ state, which is defined as
\begin{align}
\label{eq:perfect-W}
    |W_{p,N}\rangle  = &\frac{1}{\sqrt{2}}|1,0,...,0\rangle+ \frac{1}{\sqrt{2(N-1)}}\biggl[|0,1,...,0\rangle \\& +...+|0,...,1,0\rangle +|0,...,0,1\rangle\biggr], \nonumber
\end{align}
which is used for `perfect teleportation' and superdense coding \cite{agrawal_perfect_2006}. An ideal system for the synthesis of $W$ states would be easily configurable for any $W$ state type.

Regarding photonic $W$ states, the form of entanglement is highly dependent on the method used for synthesis. Bulk linear optics systems have been widely explored, resulting in $W$ states where $N$-photons are entangled by their horizontal or vertical polarization \cite{yamamoto_polarization-entangled_2002,eibl_experimental_2004}. More recently, varying approaches propose encoding the entanglement spatially \cite{grafe_-chip_2014,zheng_unified_2022, swain_generation_2023,bao_very-large-scale_2023} or sequentially (time-bin) \cite{besse_realizing_2020}, either leveraging coupling between adjacent waveguides \cite{grafe_-chip_2014, swain_generation_2023, bao_very-large-scale_2023}, adiabatic passage in $\Lambda$-type energy configurations \cite{zheng_unified_2022}, or the sequential emission of photons from a superconducting circuit \cite{besse_realizing_2020}. These demonstrations result in `single-photon' $W$ states, where the state describes the coherent superposition of a single photon across many separated modes.

Encoding the photon's entanglement with frequency is attractive due to compatibility with fiber transmission and resistance to decoherence in noisy channels \cite{xiao_efficient_2008,antonelli_sudden_2011}. Furthermore, frequency entangled states are easily converted to spatial entanglement using frequency filters. Generating frequency-entangled (or frequency-bin) $W$ states requires a nonlinear system that can access multiple optical modes. Pioneering proposals for synthesis of $W$ states exploit the nonlinearities of spontaneous four-wave mixing and single-photon detection \cite{menotti_generation_2016,fang_three-photon_2019,banic_integrated_2024}, where the entangled photons are encoded by their red or blue shift with respect to a central pump frequency. These proposals have achieved three-photon $W$ states with a $\sim10^3-10^4$ Hz estimated generation rate \cite{banic_integrated_2024}.

Here, we show how frequency-entangled, single-photon $W$ states of selective dimension and type (e.g., perfect $W$ states) can be synthesized by utilizing the quantum dynamics of optomechanical systems that exhibit forward Brillouin interactions. The proposed method is also easily altered for quantum frequency translation, which is the process of transferring photonic quantum states between different spectral modes \cite{mcguinness_theory_2011}. In contrast with backward Brillouin scattering \cite{behunin_harnessing_2023}, forward Brillouin scattering (FBS) involves a ladder of optical resonances, where scattering between any two adjacent modes is mediated by a single phonon mode \cite{kharel_noise_2016}. These dynamics are enabled in tightly confined systems (waveguides or resonators), where the system geometry dictates the frequency ($\sim$MHz-GHz) of the slow group velocity mechanical modes, and were recently examined for quantum state synthesis for the first time \cite{shepherd_multi-phonon_2024}. 

\begin{figure}
    \centering
    \includegraphics[width=8.6cm]{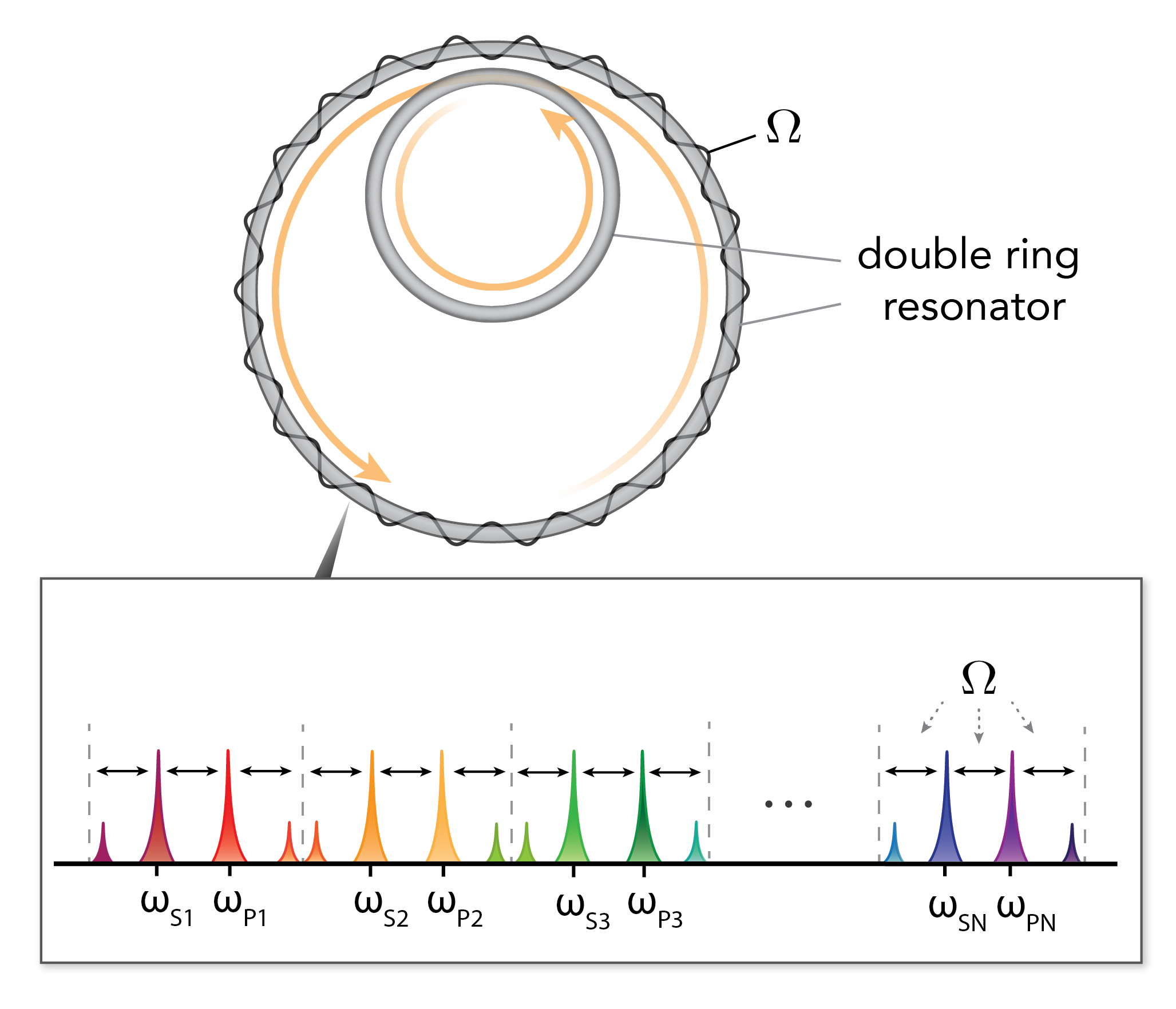}
    \caption{Example system that demonstrates forward Brillouin scattering where a double ring resonator provides periodic suppression of forward Brillouin scattering on every third optical mode, which are resonantly spaced by the phonon frequency, $\Omega$. Each pair of optical modes is labeled by $\omega_{pn}$ or $\omega_{sn}$ for each resonant pump/Stokes mode.}
   \label{fig:resonator}
\end{figure}

For our proposed methods, the FBS system will require periodic suppression of scattering, where every third optical mode is suppressed, creating an array of isolated pump-Stokes systems that all interact with the same phonon. One example system that could achieve this is shown in the form of a double ring resonator in Fig. \ref{fig:resonator}. The phonon frequency, $\Omega$, is an integer multiple of the large resonators free spectral range in absence of the smaller ring. The incorporation of the smaller ring introduces destructive interference periodically to every third resonance, creating hybridization that splits the degenerate mode at $3\Omega$ when critical coupling is achieved \cite{liu_integrated_2024, liu_tunable_2024}. The forward Brillouin-active phonon mode is shown in Fig. \ref{fig:dispersion}(a) as the white circle, and the phase matching for a candidate system with the desired suppression is shown in Fig. \ref{fig:dispersion}(b), where the stop bands create the isolated pump-Stokes systems, all resonant with $\Omega$.

For $W$ state synthesis, our methods utilize what we term a `super $\pi$-pulse'. By initially manipulating the system into a single phonon Fock state, the `super $\pi$-pulse' is accomplished by injecting simultaneous pulses on $N$-Stokes frequencies, which induces a state-swap from the single phonon to an $N$-dimensional $W$ state. The $W$ state describes a single photon existing across $N$-pump frequencies at a probability related to the corresponding Stokes pulse amplitudes. The $W$ state synthesis time and type are easily modified by adjusting the pulse amplitudes. In addition, the dimension and output frequencies of the $W$ state can be selected. 

Quantum frequency translation, commonly accomplished with spontaneous four-wave mixing \cite{mcguinness_quantum_2010, mcguinness_theory_2011}, is possible in this FBS system by addressing an injected photonic quantum state with a series of optomechanical $\pi$-pulses of different Stokes frequencies. The first pulse transfers the state into the phononic domain, and the second transfers it back to a photonic state with a different frequency. Due to long phonon lifetimes at cryogenic temperatures \cite{goryachev_extremely_2012, maccabe_nano-acoustic_2020}, the system can be held in the mechanical state for extended times, which may enable unique storage and delay capabilities.

\section{forward Brillouin Scattering}
Brillouin scattering is the inelastic scattering of incident light upon interaction with acoustic phonons. In its most commonly known form, backward Brillouin scattering, two counter-propagating light waves couple to a travelling acoustic wave \cite{boyd_chapter_2008}. Backward Brillouin dynamics have been explored in the quantum regime for state synthesis \cite{behunin_harnessing_2023}, and while showing similarities to traditional bipartite optomechanical systems, the increased complexity of the system offers greater access to exotic quantum states. FBS contrasts sharply with backward Brillouin, exhibiting co-propagating optical modes couple to a phonon mode through electrostriction and radiation pressure \cite{kharel_noise_2016}. In highly confined systems, a form of FBS (termed intra-modal) is allowed in which a cut-off mechanical mode, of frequency $\Omega$, can mediate scattering between a ladder of resonant optical frequencies \cite{shelby_resolved_1985}. Scattering to distant modes in the optical ladder is physically limited by dispersion; however, this is negligible for most candidate systems over narrow FBS frequency ranges. A variety of systems can support FBS \cite{eggleton_brillouin_2019}, including optical fibers \cite{shelby_resolved_1985, behunin_spontaneous_2019, kang_tightly_2009, kang_all-optical_2010, renninger_forward_2016}, resonator systems \cite{zhang_forward_2017, bahl_brillouin_2013}, silicon waveguides \cite{shin_control_2015, shin_tailorable_2013, kittlaus_large_2016} and optical microspheres \cite{bahl_stimulated_2011, yu_investigation_2022}. Leveraging an all fiber system for state synthesis could provide a simple route to implementing these systems at cryogenic temperatures.
\begin{figure}
    \centering
    \includegraphics[width=8.6cm]{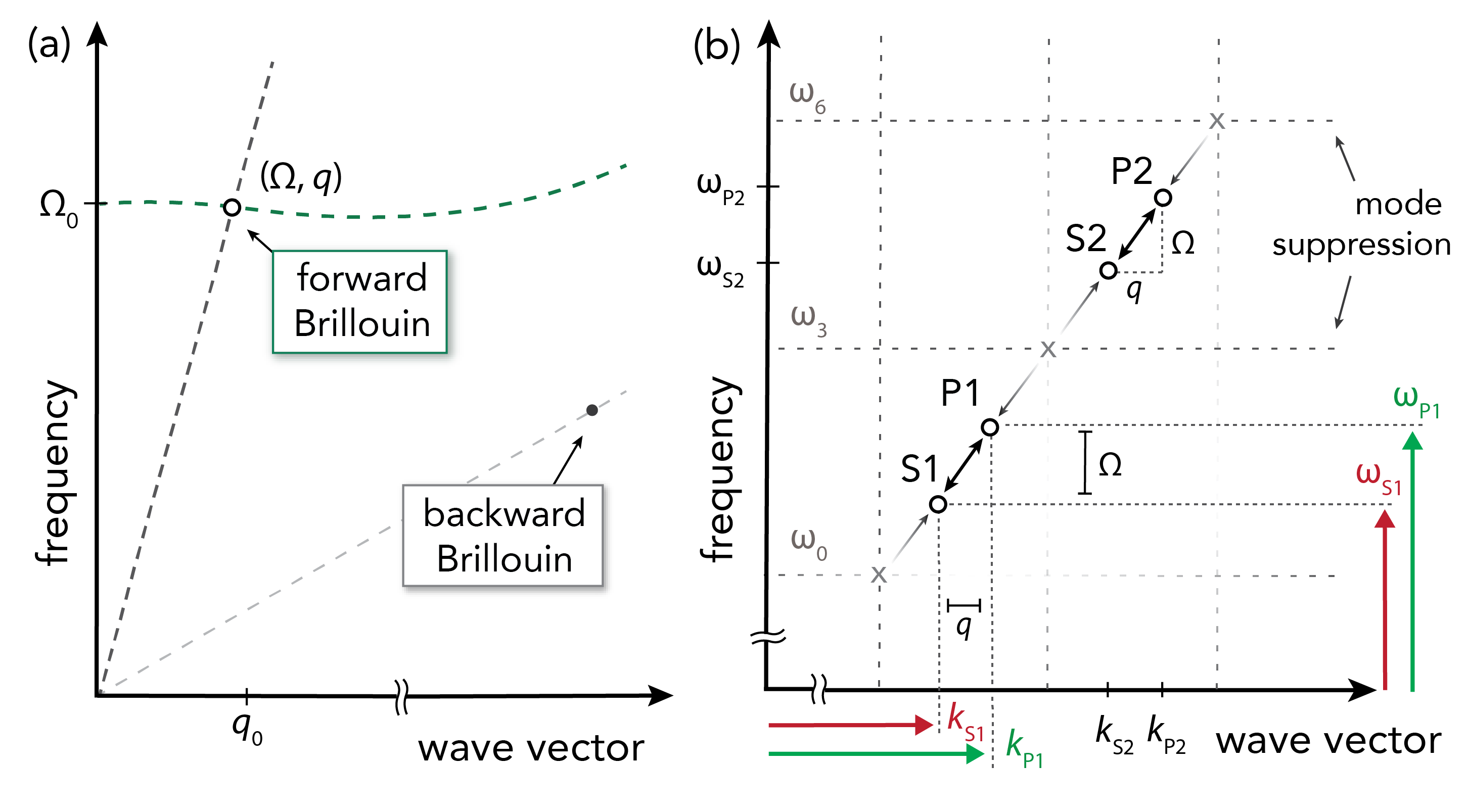}
    \caption{Phase matching for intra-modal Brillouin scattering. (a) Acoustic dispersion, where the forward Brillouin active mode is denoted as the white circle on the green line, where the mechanical frequencies $\Omega$ are approximately equal to the cut-off frequencies $\Omega_0$. This is contrasted with a backward Brillouin mode (black dot on gray line) (b) Optical dispersion, including periodic mode suppression, which create the array of pump/Stokes systems that are all resonantly coupled to the same phonon.}
   \label{fig:dispersion}
\end{figure}

The efficiency of FBS is maximized when energy and momentum are conserved, i.e., phase-matching conditions are satisfied, 
\begin{align}
\label{eq:phase-matching-1}
    \omega_m = \omega_{m-1} + \Omega \\
\label{eq:phase-matching-2}
    k_m = k_{m-1} + q.
\end{align}
Here, the ladder of optical resonances are labeled as $\omega_m$, where the integer $m$ labels the optical mode. The wavevectors $q$ and $k_m$ relate to the phonon and optical modes respectively. The optical dispersion relations of $\omega_m = c k_m/n_m$, where $n_m$ is the effective refractive index at $\omega_m$, $c$ is the speed of light, and $k_{m-1} \approx k_m - (\omega_m - \omega_{m-1}) \partial k_m/\partial \omega$ restrict Eqs. \eqref{eq:phase-matching-1} and \eqref{eq:phase-matching-2} to a required phonon phase velocity that is equal to the optical group velocity. Despite phonon phase velocities being much less than the group velocities of light in bulk materials, confined systems can support a high phase velocity of sound, with cut-off mechanical modes of frequency $\Omega\neq0$ when $q\approx0$. This is demonstrated in Fig. \ref{fig:dispersion}(a), where the chosen forward Brillouin active mode is shown as the white circle on the green phonon dispersion curve, contrasted with other Brillouin active modes as black dots on the gray lines.

FBS can be modeled with the following Hamiltonian, $H = H_0 + H_{int}$, where
\begin{align}
\label{eq:H0}
    H_0 = \displaystyle\sum^\infty_{m=-\infty}\hbar (\omega_p + m\Omega)a^{\dag}_ma_m + \hbar \Omega b^{\dag}b
\end{align}
\begin{align}
\label{eq:Hint}
  H_{int} = \hbar g \displaystyle\sum^\infty_{m=-\infty} a_m a^{\dag}_{m-1}b^{\dag} + {\rm H.c.}   
\end{align}
Equation \eqref{eq:H0} describes the free evolution of the system, and Eq. \eqref{eq:Hint} describes the interactions between adjacent optical modes and the phonon, where H.c stands for the Hermitian conjugate. The interactions are quantified by coupling strength, $g$, which is taken as a constant. Here, $\Omega$, $b$ and $b^\dag$, are the angular frequency, annihilation operator and creation operator of the phonon mode respectively. With central `pump' frequency, $\omega_p$, the first term in Eq. \eqref{eq:H0} describes the resonant optical ladder, with $a_m$ as the annihilation operator for the $m$th optical mode with angular frequency $\omega_m$, and $[a_m,a_{m'}^\dag]=\delta_{m,m'}$. 

For the synthesis of $W$ states, this FBS system will require periodic suppression of scattering on every third optical mode. This is shown in Fig. \ref{fig:resonator}, where the suppression is provided in the form of destructive interference by the second smaller resonator. This will truncate the Hamiltonian, leaving an array of pump/Stokes systems interacting with the same phonon. In addition to a double ring resonator, the required mode engineering can be realized in a variety of ways for varying FBS capable systems, such as altering the resonator structure \cite{wang_taming_2024, liu_tunable_2024, liu_integrated_2024} or using fiber Bragg gratings in a linear waveguide \cite{lee_suppression_2003, merklein_enhancing_2015, ibsen_sinc-sampled_1998}. With occupation of suppressed modes forbidden, every third term from Eqs. \eqref{eq:H0} and \eqref{eq:Hint} is removed. The truncated Hamiltonian becomes
\begin{align}
\label{eq:H0-stopbands}
    H_0 = \sum^\infty_{n=-\infty}\biggl( \hbar \omega_{pn} a_{pn}^\dag a_{pn} + \hbar \omega_{sn} a_{sn}^\dag a_{sn} \biggr) + \hbar \Omega b^\dag b
\end{align}
\begin{align}
\label{eq:Hint-stopbands}
    H_{int} = \sum^\infty_{n=-\infty}\hbar g \biggl( a_{pn} a_{sn}^\dag b^\dag + a_{pn}^\dag a_{sn}b \biggr)
\end{align}
where modes $m=0 + 3n\Omega$ are suppressed, with integer $n=-\infty,...,\infty$. Modes $m=2 + 3n\Omega$ and $m=1 + 3n\Omega$ have been replaced with $pn$ and $sn$ for each isolated pump/Stokes system, indexed by $n$. With this change, the phase matching conditions can be written as 
\begin{align}
\label{eq:phase-matching-stopband-1}
    \omega_{pn} = \omega_{sn} + \Omega \\
\label{eq:phase-matching-stopband-2}
    k_{pn} = k_{sn} + q,
\end{align}
which are shown with optical dispersion in Fig. \ref{fig:dispersion}(b). It is important to mention that Fig. \ref{fig:dispersion}(b) holds for linear waveguide systems. A resonator supports a discrete set of wavevectors compared to a continuous line.
\begin{figure}
    \centering
    
    \includegraphics[width=8.6cm]{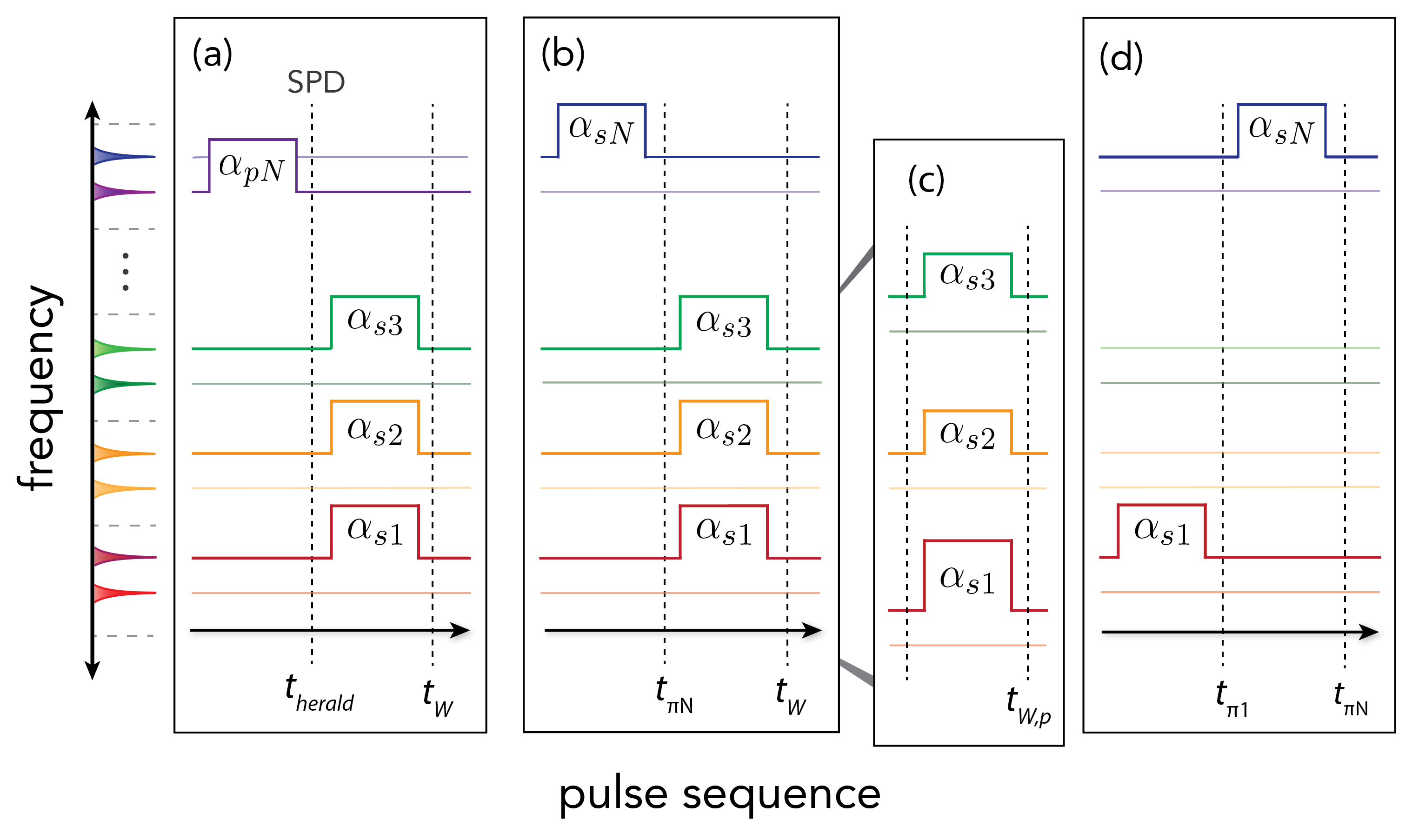}
    \caption{Pulse sequencing for (a) a standard $W$ state where a heralding event (denoted by SPD for single photon detection) projects the system to a single phonon Fock state, (b) a standard $W$ state where a single photon is injected and swapped to a single phonon. (c) shows how both (a) and (b) can be altered for a perfect $W$ state by adjusting the pulse amplitudes, represented here by the height difference. (d) shows pulse sequencing for quantum frequency translation.}
   \label{Fig:pulse_sequence}
\end{figure}

\section{quantum protocols}
For a theoretical description of all quantum protocols described here, we examine time dependent solutions to the Schrodinger equation, which are given by application of the time evolution operator, $U=\exp(-iHt/\hbar)$, to an initial state, $|\psi(0)\rangle$, such that $|\psi(t)\rangle = U(t,0)|\psi(0)\rangle$. When the phase-matching conditions in Eqs. \eqref{eq:phase-matching-stopband-1} and \eqref{eq:phase-matching-stopband-2} are satisfied, the interactions described by Eq. \eqref{eq:Hint-stopbands} conserve energy, i.e., $[H,H_{int}]=0$, allowing the time evolution operator to be factorized into the form
\begin{equation}
    U(t) = U_0 U_{int} = \exp(-iH_0t/\hbar) \times \exp(-iH_{int}t/\hbar).
\end{equation}
As $U_0|\psi(0)\rangle$ does not affect global probabilities, all results moving forward will be in the interaction picture (i.e., $|\psi(t)\rangle = U_{int}|\psi(0)\rangle$). For analytical simplicity, this method assumes no decoherence; however, simulated results with optical losses will follow. Mechanical decoherence is neglected throughout as the interaction times can be made much shorter than state of the art phonon coherence times \cite{k_j_satzinger_quantum_2018, maccabe_nano-acoustic_2020}. 

{\it $W$ state synthesis}: 
By taking the phonon to initially occupy the quantum ground state via cryogenic cooling, the system must first be manipulated into a single phonon Fock state. In traditional optomechanical systems, this can be accomplished in two ways: (1) pairing two-mode squeezing with single photon detection \cite{galland_heralded_2014, hong_hanbury_2017} (see Appendix A), or (2) state-swapping an injected photon with an optomechanical $\pi$-pulse via the beamsplitter transformation \cite{rakhubovsky_photon-phonon-photon_2017} (see Appendix B). With the two-mode squeezer transformation given by $S(\xi)=\exp(\xi a^\dag b^\dag - \xi^* ab)$, and the beamsplitter transformation as $B(\mu)=\exp(\mu a^\dag b - \mu^* a b^\dag)$ \cite{aspelmeyer_cavity_2014}, we will show that although this FBS system includes more optical modes than traditional optomechanical systems, the dynamics can be reduced for specific initial conditions to match the dynamics of traditional optomechanics where single phonon Fock states have been achieved.

Benefiting from the engineered mode suppression, any isolated pump/Stokes system in which both modes are in vacuum will not interact with the rest of the system, following from $a_{pj}a_{sj}^\dag b^\dag|0_{jp},0_{sj},N_{ph}\rangle=0$ and $a_{pj}^\dag a_{sj} b|0_{pj},0_{sj},N_{ph}\rangle=0$. By considering driving light in only one pump and/or adjacent Stokes mode, we can reduce the interaction Hamiltonian to only three modes,
\begin{align}
\label{eq:Hint3mode}
    H_{int,reduced}=\hbar g(a_{p}a_{s}^\dag b^\dag+a_{p}^\dag a_{s} b),
\end{align}
which now matches that of backward Brillouin scattering \cite{behunin_harnessing_2023}.

When injecting strong laser pulses, the Hamiltonian is well-approximated by replacing the creation and annihilation operators with the coherent state's complex amplitude (i.e., $a \simeq \alpha$ and $a^\dag \simeq \alpha^*$ when $|\alpha|\gg1$). This means that by driving strong pulses on the pump ($|\alpha_{p}|\gg 1$), or Stokes mode ($|\alpha_{s}| \gg 1$), the time evolution operator of this system matches either the two-mode squeezer transformation, where $\xi=-igt\alpha_{p}$, or the beamsplitter transformation where $\mu=-igt\alpha_{s}$; therefore, the methods for single phonon preparation by \cite{galland_heralded_2014, hong_hanbury_2017, rakhubovsky_photon-phonon-photon_2017} (described in Appendices A and B), are possible in this system. It is important to note that in either case, the system can be held in the single phonon state for extended times, leveraging long phonon lifetimes at cryogenic temperatures. This allows for flexibility as to when the synthesis of the $W$ state must occur; therefore, enabling unique delay capabilities.

Once a single phonon is obtained, laser pulses on $N$-Stokes modes, with frequencies $\omega_{sn}$, are injected into the system simultaneously. This combination of pulses can realize a `super $\pi$-pulse', where the single phonon is state-swapped to a $W$ state. This process is shown in Fig. \ref{Fig:pulse_sequence}, where Fig. \ref{Fig:pulse_sequence}(a) shows the case of heralding a single phonon and Fig. \ref{Fig:pulse_sequence}(b) shows the case of injecting a single photon and swapping it to a single phonon. 

We once again leverage the strong laser approximation by considering strong lasers driven on $N$-Stokes modes. With this approximation, $U_{int}$ becomes
\begin{equation}
\label{eq:superbeamsplitter}
    U_{int} \simeq e^{-igt(A b^\dag + A^\dag b)},
\end{equation}
with
\begin{align}
    A = \displaystyle\sum^N_{n=1} a_{pn} \alpha^{*}_{sn} \quad {\rm and} \quad A^\dag = \displaystyle\sum^N_{n=1} a^{\dag}_{pn} \alpha_{sn}.
\end{align}
Here, the sum includes only the $N$-relevant pump/Stokes systems. Equation \eqref{eq:superbeamsplitter} acts like a `super beamsplitter', allowing state transfer between the phonon mode and the collection of pump modes. By making the strong laser approximation, we factorize $U_{int}$ using Wei-Norman methods (see Appendix D). After factorization (with  $\alpha_{sn} = r_n e^{i\phi_n}$, $n$ indexing each Stokes mode, $r_n=|\alpha_{sn}|$, and $\phi_n=$ arg($\alpha_{sn}$)), the interaction time evolution operator becomes
\begin{align}
\label{eq:TEO}
    U_{int} = e^{X(t)A^\dag b}e^{Y(t) \hat{\Theta}}e^{Z(t)Ab^\dag},
\end{align}
with
\begin{align}
    X(t) = Z(t) = \frac{-i \tan\bigl(gt \sqrt{\eta}\bigr)}{\sqrt{\eta}} 
\end{align}
\begin{align}
    Y(t) = \frac{-\ln \cos \bigl(gt\sqrt{\sum^N_{n=1}r_n^2}\bigr)}{\eta},
\end{align}
and
\begin{align}
    \label{eq:sum_of_squared_amp}
    \eta = \sum^N_{n=1}r_n^2.
\end{align}
The operator $\hat{\Theta} = [Ab^\dag,A^\dag b]$, which is easily computed for finite $N$. As all Stokes modes are treated as $c$-numbers, the operators $\hat{A}$, $\hat{B}$, and $\hat{\Theta}$ only act on the collection of pump modes and the phonon. For simplicity, the $s$ and $p$ will be dropped from indexed subscripts. Any $\alpha_n=r_ne^{i\phi_n}$ will refer to strong lasers on a Stokes frequency of mode $n$, and any kets labeled with $n$ will refer to a pump frequency of mode $n$. 

We now apply $U_{int}$, acting as the `super beamsplitter', to an initial state for the time-dependent wavefunction. With our intial state, $|\psi(0)\rangle = |vac\rangle_{opt}|1\rangle_{ph}$, where `opt' describes the collection of pump modes, $U_{int}|\psi(0)\rangle$ gives the time dependent wavefunction,
\begin{align}
\label{eq:wavefunction1}
    |\psi(t)\rangle = & \cos\bigl(gt\sqrt{\eta}\bigr)|vac\rangle_{opt}|1\rangle_{ph} \\&- i\sin\bigl(gt\sqrt{\eta}\bigr)|\varphi\rangle_{opt}|0\rangle_{ph}, \nonumber
\end{align}
where we've introduced a compact notation for the general $W$ state,
\begin{align}
    \label{eq:general_w}
    |\varphi\rangle=\frac{1}{\sqrt{\eta}}&\bigl[r_1e^{i\phi_1}|1,0,...,0\rangle + r_2e^{i\phi_2}|0,1,...,0\rangle \\&+... + r_Ne^{i\phi_N}|0,...,0,1\rangle \bigr] \nonumber
\end{align}
in which each value in the kets contained in $|\varphi\rangle$, separated by commas, represent each pump mode from $n=1$ to $N$. Here, the coefficients of each ket are easily modifiable by the pulse amplitude and phase.
\begin{figure}
    \centering
    
    \includegraphics[width=8.6cm]{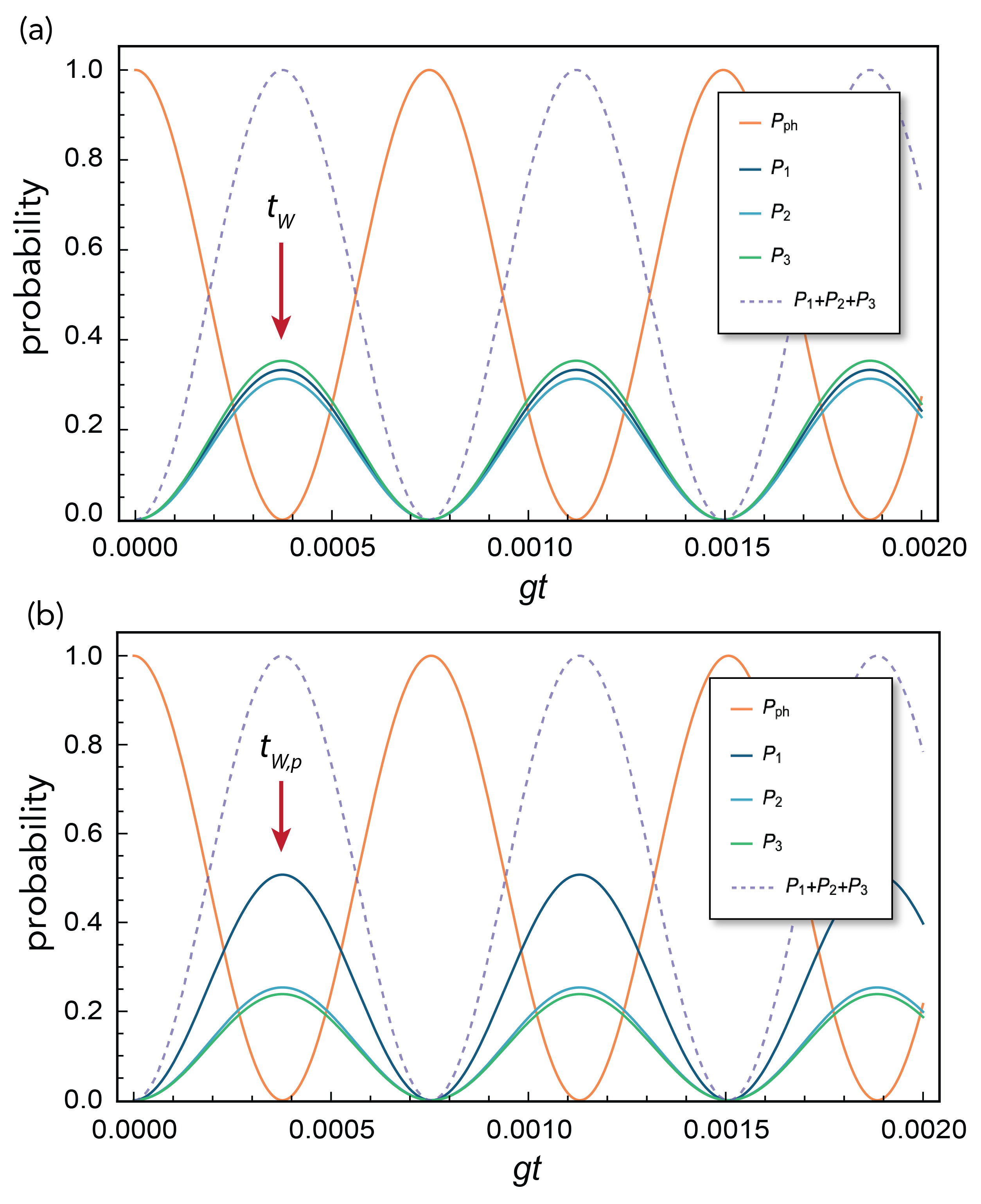}
    \caption{Probabilities of states plotted over $gt$, when the system evolves from a single phonon. (a) shows evolution to a standard $N=3$ $W$ states, with $\alpha=2424$. (b) shows evolution to a perfect $N=3$ $W$ state, with $\alpha=2100$, chosen to match the period of (a). There is a slight difference in $r_n$'s simply so they can be seen on the plot. $W$ states are denoted by red arrows. The dotted line, being the sum of the photonic states probabilities, represents the $W$ state through time.}
   \label{Fig:protocol-1-probability}
\end{figure}
Turning $|\varphi\rangle$ into a standard $W$ state, matching Eq. \eqref{eq:standard-W-state}, requires setting all $r_n$ equal. Letting $r_n=\alpha$ and setting all phases $\phi_n=0$, simplifies the wave function to
\begin{align}
    |\psi(t)\rangle =& \cos(g\alpha t \sqrt{N})|vac\rangle_{opt}|1\rangle_{ph} \\&-i\sin(g\alpha t \sqrt{N})|W_N\rangle|0\rangle_{ph}. \nonumber
\end{align}
A state-swap from phonon to $W$ state occurs, achieved with the `super $\pi$-pulse', occurs at
\begin{equation}
    t_{W} = \frac{\pi}{2g\alpha \sqrt{N}}.
\end{equation}
where $\alpha$ is the adjustable pulse amplitude, set to be constant for all frequencies, and $N$ is the $W$ state dimension, which is determined by the number of driven Stokes modes (driving on only two modes enables Bell-states of type $|\Psi\rangle$). Probabilities of each state are plotted in Fig. \ref{Fig:protocol-1-probability}(a) for $N=3$ over $gt$. Each state containing a single photon of frequency $\omega_{pn}$, and vacuum in the phonon, exists with probability $P_n$. The state containing the single phonon and vacuum in the optical modes is labeled $P_{ph}$. The photonic states have slightly different probabilities because slightly different $\alpha$'s were used--only for the purpose of clarity on the plot. The dashed line represents the $W$ state, which is Rabi-oscillating with the phonon. It is important to mention that after the `super $\pi$-pulse', the $W$ state will oscillate slowly back to the single phonon state; however, cavity decay times are easily engineered to be $\ll \pi/2g$ (following from the vacuum Rabi-frequency, $gt$), thus the $W$ state will be emitted long before this occurs.

In the same way $|\varphi\rangle$ was manipulated into $|W_N\rangle$ by adjusting the pulse amplitudes to be equal, it can also be easily manipulated into a perfect $W$ state with specifically chosen pulse amplitudes (see Fig. \ref{Fig:pulse_sequence}(c)). The perfect $W$ state, $|W_{p,N}\rangle$, is defined in Eq. \eqref{eq:perfect-W}. To appropriately adjust the laser amplitudes for a perfect $W$ state of $N$-dimension, set all phases to zero and
\begin{align}
    \frac{r_1}{\sqrt{\eta}} = \frac{1}{\sqrt{2}} \quad {\rm and} \quad \frac{r_n}{\sqrt{\eta}} = \frac{1}{\sqrt{2(N-1)}}.
\end{align}
Let $r_n=\alpha$ for each $n$ from $2$ to $N$, as those coefficients are all equal in $|W_{p,N}\rangle$. Solving for the ratio of the laser amplitudes gives
\begin{equation}
    \frac{r_1}{\alpha}=\sqrt{N-1}.
\end{equation}
The time for a state-swap from a phonon to an $N$-dimensional perfect $W$ state is then
\begin{equation}
    t_{W,p} = \frac{\pi}{2g\alpha\sqrt{2(N-1)}}
\end{equation}
Figure \ref{Fig:protocol-1-probability}(b) shows how the probabilities of each state differ for an $N=3$ perfect $W$ state, where the laser amplitudes have been chosen to give the same Rabi-frequency as Fig. \ref{Fig:protocol-1-probability}.

{\it Quantum frequency translation}: 
This system can facilitate quantum frequency translation, in which there is a coherent transfer of a quantum state between optical modes of distict frequencies. Here, a series of $\pi$-pulses (shown in Fig. \ref{Fig:pulse_sequence}(d)) accomplishes this by swapping into and then out of the phononic domain. An injected pump photon of frequency $\omega_{pi}$, is easily swapped to a single phonon with a strong laser, $|\alpha_{si}|\gg1$, inducing an optomechanical $\pi$-pulse via the beamsplitter transformation, $B(\mu)=\exp(\mu a b^\dag + \mu^* a^\dag b)$ (see Appendix B), such that
\begin{align}
    B_i(\pi/2)|1_i,0_j,0_{ph}\rangle = -e^{i\phi_i}|0_i,0_j,1_{ph}\rangle,
\end{align}
where $|\mu| = g|\alpha_{i}|t=\pi/2$. With $\alpha_{sn}=r_ne^{i\phi_n}$, we define the state-swap time for a $\pi$-pulse on general mode $\omega_{sn}$ as
\begin{align}
\label{eq:pi}
    t_{\pi n} = \frac{\pi}{2gr_n}
\end{align}
Applying a second $\pi$-pulse with $|\alpha_{sj}|\gg1$, swaps the single phonon to a single photon of a new frequency $\omega_{pj}$,
\begin{align}
   B_j(\pi/2)(-e^{i\phi_i})&|0_i,0_j,1_{ph}\rangle = \\&-e^{i(\phi_i-\phi_j)}|0_i,1_j,0_{ph}\rangle, \nonumber
\end{align}
at $t_{\pi i}+t_{\pi j}$, carrying only the phase factors of the two Stokes pulses. This method also works for any general photonic state as well. Consider injecting the state $|\Phi\rangle=\sum_k C_k |k\rangle$ with frequency $\omega_{pi}$, which could describe a coherent state, squeezed light, or a superposition of Fock states. Once injected into the system, the initial state can be written as
\begin{align}
    |\psi(0)\rangle = |\Phi_i,0_j,0_{ph}\rangle = \sum_k C_k \frac{(a_i^\dag)^k}{\sqrt{k!}} |0_i,0_j,0_{ph}\rangle.
\end{align}
Noting that $B(a^\dag)^k B^\dag = (Ba^\dag B^\dag)^k$, a $\pi$-pulse with $|\alpha_{si}|\gg1$ gives
\begin{align}
    B_i(\pi/2)|\Phi_i,0_j,0_{ph}\rangle = -e^{(i\phi_i)^k}|0_i,0_j,\Phi_{ph}\rangle,
\end{align}
at $t_{\pi i}$. Applying the second $\pi$-pulse on $\omega_{pj}$ gives
\begin{align}
    B_j(\pi/2)(-e^{ik\phi_i})&|0_i,0_j,\Phi_{ph}\rangle = \\&-e^{ik(\phi_i-\phi_j)}|0_i,\Phi_j,0_{ph}\rangle\nonumber
\end{align}
at $t_{\pi i}+t_{\pi j}$, and thus the injected general photonic state $|\Phi\rangle$ has been converted in frequency from $\omega_{pi}$ to $\omega_{pj}$. In both cases, the swap time is the same, making quantum frequency translation in this FBS system occur at
\begin{align}
    \label{eq:QFT-time}
    t_{qft}=t_{\pi i} + t_{\pi j} = \frac{\pi}{2g}\biggl(\frac{1}{r_i}+\frac{1}{r_j}\biggr).
\end{align}

These systems are highly tailorable via design and geometry to accommodate varying injected photon frequencies, making it a logical candidate for integration with other systems. In addition, this system enables highly selective output frequencies ranging over tens of GHz, at tailorable GHz intervals.

{\it Fidelity of $W$ states}:
We examine the dynamics in the presence of optical loss, $\gamma$, on all pump modes in order to study the effects of decoherence. Assuming no phonon decoherence, and a bath temperature of zero due to the high optical frequencies, we simulate the Lindblad master equation,
\begin{equation}
    \dot{\rho}=-\frac{i}{\hbar}[H,\rho]+\frac{\gamma}{2} \sum_j[2a_{pj}\rho a_{pj}^\dag -a_{pj}^\dag a_{pj}\rho - \rho a_{pj}^\dag a_{pj}],
\end{equation}
to determine the fidelity of our $W$ state synthesis and quantum frequency translation protocols. For highest fidelity, the pulse amplitudes should be as large as possible regarding available laser power, while also considering limiting excess heating of the phonon mode \cite{doeleman_brillouin_2023}. We define a maximum laser amplitude $\alpha_{max}=\sqrt{\eta}$, which in the case of a resonator, corresponds to the intra-cavity photon number, $|\alpha_{max}|^2=n_{cav}$, and thus also the circulating power $P=\hbar \omega v_g n_{cav}/L$. By introducing $\alpha_{max}$, we can redefine the optomechanical $\pi$ (or super $\pi$)-pulse time as $\tau(\alpha_{max})$, which is constant across all protocols, and any $W$ state dimension, for a given maximum amplitude,
\begin{align}
    \tau(\alpha_{max})=\frac{\pi}{2g\alpha_{max}}.
\end{align}
Using $g\sim(2\pi)15$ kHz and $\gamma/g\simeq1800$, estimated from physically realized systems \cite{kittlaus_large_2016, shin_tailorable_2013, zhang_forward_2017}, an optomechanical $\pi$ (or super $\pi$)-pulse with a fidelity $>70\%$ requires $P\simeq50$ mW and $L\simeq 4$ mm ($\alpha_{max}=4200$), and in turn gives a $\pi$-pulse duration of $\tau\sim4$ ns. 

In the case of an injected photon, for either $W$ state synthesis or quantum frequency translation, the fidelity drops to $\simeq52\%$ as the total interaction time doubles. State of the art systems may soon reach the regime where $\gamma \sim g$ for forward Brillouin scattering \cite{maccabe_nano-acoustic_2020, liu_ultralow_2022}, and these protocols would support a fidelity near unity (e.g., an optomechanical $\pi$-pulse fidelity is $98\%$ in systems where $\gamma/g=100$ and $\alpha_{max}=4200$).

\section{Conclusion}
In conclusion, we have proposed a system that utilizes the quantum dynamics of forward Brillouin scattering for fast, high-fidelity generation of frequency-bin, single-photon $W$ states in a versatile manner. $W$ state dimension, type, output frequency, and interaction time are adjustable by selecting the pulse amplitudes. The protocols can be modified to allow quantum frequency translation, where an injected photonic state's frequency can be selectively changed. While the protocols require one single-photon detection event for preparation of the initial state (by heralding a single phonon in the system or heralding a single photon to be injected), the following dynamical evolution is deterministic. Although interaction times are on the order of a few nanoseconds, the generation rate also relies on the rate at which the initial state is prepared. Current single-photon generation rates \cite{ma_ultrabright_2020} and reported single phonon heralding rates \cite{hong_hanbury_2017} permit $W$ state generation on the order of MHz. The proposed systems' ability to accept a single photon and convert it to a variety of complex optical states demonstrates the possibility for integrating this system with others for a variety of quantum protocols. In addition to fast synthesis of high-fidelity, frequency-bin $W$ states and quantum frequency translation, this work is the first demonstration showing the versatility of forward Brillouin scattering as a tool for quantum state synthesis of light.

\section{Appendix}
\subsection{Heralding a single phonon Fock state}
\label{app:A}
The two-mode squeezer is defined as $S(\xi)=\exp(\xi a_s^\dag b^\dag - \xi^* a_sb)$. Applying this transformation to the vacuum initial state, $|\psi(0)\rangle = |0\rangle_s|0\rangle_{ph}$, gives
\begin{align}
    |\psi(t)\rangle = S(\xi)|\psi(0)\rangle=\sum_{n=0}^\infty C_n |n\rangle_s |n\rangle_{ph},
\end{align}
where the scattered Stokes photon and phonon occupation number is correlated. Pairing this result with single photon detection of a Stokes photon, $\langle 1_s |\psi(t)\rangle$, collapses the wavefunction by introducing the term $\delta_{1,n}$. Noting that the complex argument can be written as $\xi=re^{i\phi}$, accounting for the real part, $r$, and the imaginary part, $e^{i\phi}$, the detection event heralds the state $|\psi_{ph}[1_s,t)\rangle=e^{i\phi}|1\rangle_{ph}$ after normallizing.

\subsection{State-swapping a single photon to a single phonon}
The beamsplitter transformation is defined as $B(\mu)=\exp(\mu a^\dag b - \mu^* a b^\dag)$. When applied to a single pump photon, the dynamics can be written as
\begin{align}
    Ba^\dag|0\rangle_p|0\rangle_{ph} = Ba^\dag B^\dag B |0\rangle_p|0\rangle_{ph},
\end{align}
nothing that $B^\dag B=1$. Using the Baker-Campbell-Hausdorf formula, $e^{\hat{X}}\hat{Y}e^{-\hat{X}}=\hat{Y}+[\hat{X},\hat{Y}]+1/2![\hat{X},[\hat{X},\hat{Y}]]$, and the fact that $B|0\rangle_p|0\rangle_{ph}$, the state becomes 
\begin{align}
    \label{eq:beamsplitter_wavefunction}
    |\psi(t)\rangle = \biggl(a^\dag \cos(|\mu|) - \frac{\mu^*}{|\mu|}b^\dag\sin(|\mu|)\biggr)|0\rangle_p|0\rangle_{ph}.
\end{align}
Letting $\mu=|\mu|e^{i\phi}$ to account for its real and imaginary parts, reduces the term $\mu^*/|\mu|=e^{i\phi}$. Equation \eqref{eq:beamsplitter_wavefunction} shows that a state-swap, also known as a $\pi$-pulse, can be accomplished optomechanically by setting $|\mu|=\pi/2$. In the FBS system described here, $\mu=-igt\alpha_{sj}$, when driving a pulse on resonant Stokes mode $n=j$.

\subsection{Wei-Norman factorization of the `super beamsplitter'}
Following the general principles from the previous section, we can factorize the time evolution operator for a system driven on Stokes' frequencies with strong lasers. The interaction Hamiltonian becomes:
\begin{equation}
    H_{int} = \hbar g \biggl[\sum^N_{n=1} a_{pn} \alpha_{sn}^* \biggr]b^\dag + \hbar g \biggl[\sum^N_{n=1} a^\dag_{pn} \alpha_{sn} \biggr]b
\end{equation}
The time evolution operator is:
\begin{equation}
    U = e^{-igt(\hat{A}+\hat{B})}
\end{equation}
with 
\begin{equation}
    \hat{A}=\biggl[\sum^N_{n=1} a_{pn} \alpha_{sn}^* \biggr]b^\dag
\end{equation}
\begin{equation}
    \hat{B} = \biggl[\sum^N_{n=1}a^\dag_{pn} \alpha_{sn}  \biggr]b
\end{equation}
The sum is what is responsible for distinguishing what we call a 'super' beamsplitter and a standard beamsplitter. Building Lie-algebra, Let $\alpha_{sn} = r_ne^{i\phi_n}$ (note that although these are denoted the same as the previous section, they are being applied on the Stokes' modes):
\begin{equation}
    [\hat{A},\hat{B}] = \biggl[\sum^N_{n=1} a_{pn} r_n e^{-i\phi_n} b^\dag , \sum^N_{n=1} a^\dag_{pn} r_n e^{i\phi_n} b\biggr] = \hat{\Theta}
\end{equation}
This is easily solvable for a specific dimension of N using:
\begin{equation}
    [a_1 a_2^\dag , a_1^\dag a_2] = n_1-n_2
\end{equation}
Inspecting $\hat{\Theta}$'s commutator with $\hat{A}$ and $\hat{B}$ gives
\begin{equation}
    [\hat{A},\hat{\Theta}] = -2\eta\hat{A}
\end{equation}
\begin{equation}
    [\hat{B},\hat{\Theta}] = 2\eta\hat{B}
\end{equation}
The same steps for factorization occur as before, and although the commutators look very similar, the difference in sign results in a change from hyperbolic trig functions to trig:
\begin{equation}
    a(t) = \frac{-i\tan{\biggl(gt\sqrt{\eta}}\biggr)}{g\sqrt{\eta}}
\end{equation}
\begin{equation}
    b(t) = \frac{-i\tan{(gt\sqrt{\eta}})}{g\sqrt{\eta}}
\end{equation}
\begin{equation}
    \theta(t) = \frac{-\ln\cos(gt\sqrt{\eta})}{g\eta}
\end{equation}
Once again, the factored time evolution operator is:
\begin{equation}
    U_{int} = e^{a(t)\hat{A}}\times e^{\theta(t)\hat{\Theta}}\times e^{b(t)\hat{B}}
\end{equation}
but with a, b, and $\theta$ being different-- as well as $\hat{A}$,$\hat{\Theta}$, and $\hat{B}$. The easiest state to apply this to is a phonon Fock state, $|k\rangle$.
\begin{equation}
    \begin{split}
      |\psi(t)\rangle = &\cos^k\Bigl(gt\sqrt{\eta}\Bigr)\prod^N_{l=1}\sum^{A}_{m_l=0}\Bigl[\frac{-i\tan\bigl(gt\sqrt{\eta} \bigr)}{\sqrt{\eta}}\Bigr]^{m_l}\\& \times r^{m_l}_le^{-im_l\phi_l}\frac{\sqrt{k!}}{\sqrt{m_l!}\sqrt{(k-\sum^N_{j=1}m_j)!}}\\&\times|m_1,...,m_N\rangle_{opt}\otimes|k-\sum^N_{i=1}m_j\rangle_{ph}
    \end{split}
\end{equation}
with A being a placeholder for
\begin{equation}
    k-\sum_{j=1}^{l-1}m_j
\end{equation}
In the results of this paper, $k=1$. When only using 1 laser, this result also reduces to a standard optomechanical beamsplitter.

\subsection{$N$-lasers turned on prior to single photon injection}
Although an injected single photon is easily swapped to a single phonon with one pulse frequency, which can then be converted to a photonic $W$ state with multiple pulse frequencies, there may be a benefit to having all Stokes lasers turned on prior to the photon's injection. In this case, specially tuning the laser amplitudes can also achieve a $W$ state, without the need for precise pulse timing. In this approach, application of the time evolution operator in Eq. \eqref{eq:TEO} on a single photon, $U_{int}|1,0,...,0\rangle_{opt}|0\rangle_{ph}$ (choosing $n=1$ as the mode of the injected photon for simplicity), gives the wavefunction
\begin{align}
    &|\psi(t)\rangle = \frac{r_1e^{-i\phi_1}}{\sqrt{\eta}}\Bigl[\cos\bigl(gt\sqrt{\eta}\bigr)-1\Bigr]|\varphi\rangle_{opt}|0\rangle_{ph} \\& - \frac{ir_1e^{-i\phi_1}}{\sqrt{\eta}}\sin\biggl(gt\sqrt{\eta}\biggr)|vac\rangle_{opt}|1\rangle_{ph} \nonumber + |1,...,0\rangle_{opt}|0\rangle_{ph}. \nonumber
\end{align}
where once again the commas in the ket labeled `opt' distinguish pump modes from $n=1$ to $N$ and all Stokes lasers are described by $\alpha_n=r_ne^{i\phi_n}$, noting that $\alpha_1$ is specifically the mode adjacent to the injected photon in this case. $|\varphi\rangle$ is defined in Eq. \eqref{eq:general_w}. Here, the $W$ state can only exist when $\cos(gt\sqrt{\eta})=-1$, thus the W-state synthesis time is
\begin{align}
\label{eq:swap-protocol-2}
    t_W = \frac{\pi}{g\sqrt{\eta}}.
\end{align}
For either type of $W$ state, this method requires setting all $r_n=\alpha$ over modes $n=2$ to $N$, resulting in equal probability amplitudes. For a standard $W$ state, setting $P_1=P_n$ at $t_W$ reveals the ratio of amplitudes $r_1/\alpha$ that achieves $|W_N\rangle$,
\begin{align}
\label{eq:laser-ratio}
    \frac{r_1}{\alpha}= \frac{N-1}{\sqrt{N}\pm1}.
\end{align}
\begin{figure}
    \centering
    \includegraphics[width=8.6cm]{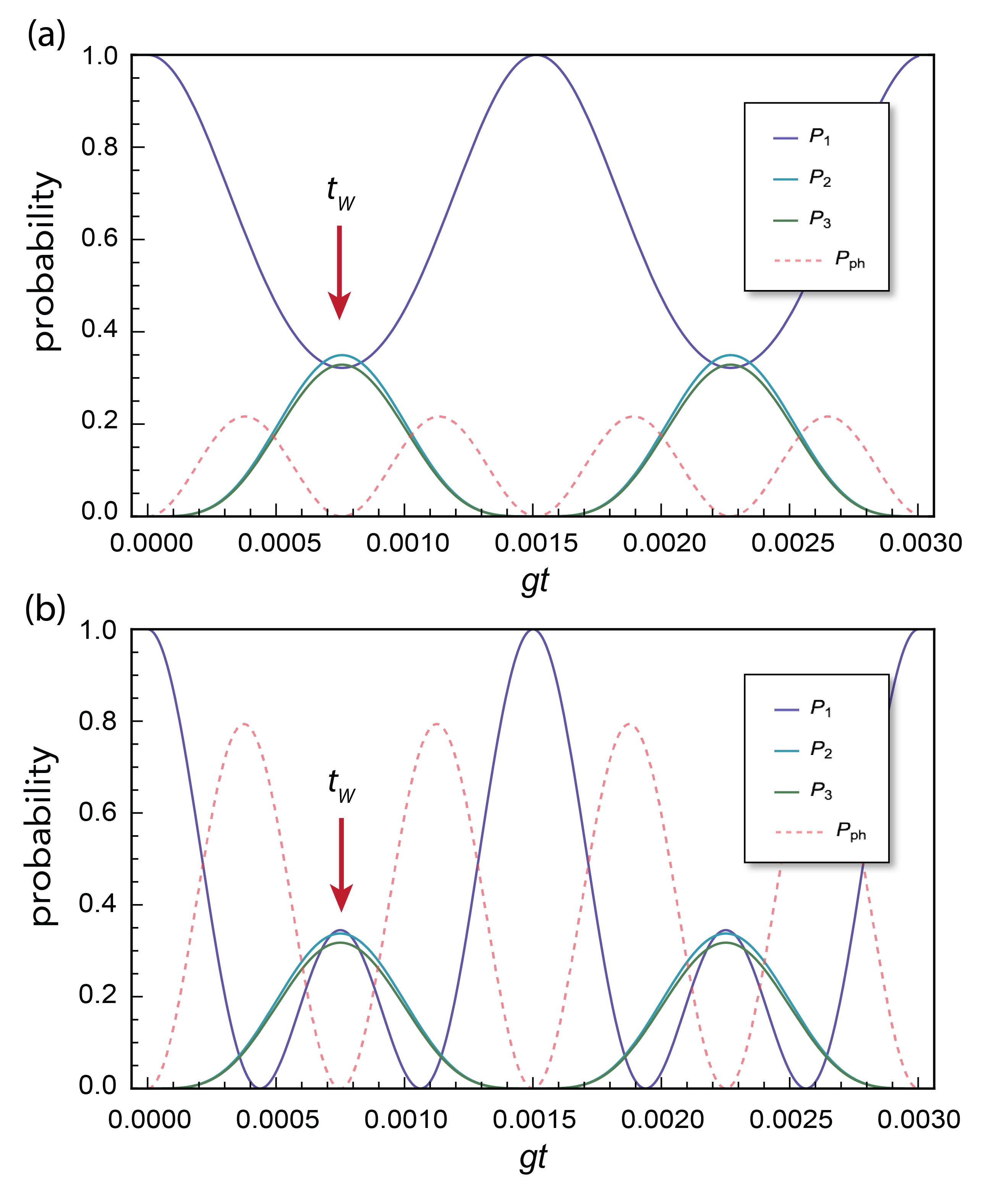}
    \caption{Probabilities of states plotted over $gt$ with injected single photon while all Stokes lasers are turned on, showing evolution to standard $N=3$ W-states. (a) corresponds to addition in Eq. \eqref{eq:laser-ratio}, with $\alpha=2637$. (b) corresponds to addition in Eq. \eqref{eq:laser-ratio}, with $\alpha=1365$. There is a slight difference in $P_2$ and $P_3$ is simply so they can be seen on the plot. W-states are denoted by red arrows. Amplitudes were chosen specifically to have the same period as Fig. \ref{Fig:protocol-1-probability}.}
   \label{Fig:classic-W-protocol-2}
\end{figure}

For an $N=3$ dimensional $W$ state, the probabilities of each ket are plotted over $gt$ in Fig. \ref{Fig:classic-W-protocol-2}. Fig. \ref{Fig:classic-W-protocol-2}(a) shows the dynamics for when Eq. \eqref{eq:laser-ratio} has subtraction in the denominator, whereas Fig. \ref{Fig:classic-W-protocol-2}(b) is the case of addition. The $W$ state occurs at the intersection of the three solid lines, denoted with a red arrow.

Creating a perfect $W$ state of arbitrary dimension (described in Eq. \eqref{eq:perfect-W}) once again requires specially adjusting the pulse amplitudes; however, in this case the laser amplitudes should be chosen so that state $|1,0,...,0\rangle_{opt}|0\rangle_{ph}$, with probability $P_1$, is equal to $1/2$ at $t_W$ from Eq. \eqref{eq:swap-protocol-2}, and all other states with probability $P_n$ equal to $1/(2(N-1))$ at $t_W$. Setting $r_n=\alpha$ over modes $n=2$ to $N$,  the ratio of $r_1/\alpha$ for a perfect W-state is
\begin{equation}
\label{eq:ratio-perfect-W}
    \frac{r_1}{\alpha}=\sqrt{\frac{N-1}{3\pm\sqrt{8}}}.
\end{equation}


\begin{thebibliography}{62}%
\makeatletter
\providecommand \@ifxundefined [1]{%
 \@ifx{#1\undefined}
}%
\providecommand \@ifnum [1]{%
 \ifnum #1\expandafter \@firstoftwo
 \else \expandafter \@secondoftwo
 \fi
}%
\providecommand \@ifx [1]{%
 \ifx #1\expandafter \@firstoftwo
 \else \expandafter \@secondoftwo
 \fi
}%
\providecommand \natexlab [1]{#1}%
\providecommand \enquote  [1]{``#1''}%
\providecommand \bibnamefont  [1]{#1}%
\providecommand \bibfnamefont [1]{#1}%
\providecommand \citenamefont [1]{#1}%
\providecommand \href@noop [0]{\@secondoftwo}%
\providecommand \href [0]{\begingroup \@sanitize@url \@href}%
\providecommand \@href[1]{\@@startlink{#1}\@@href}%
\providecommand \@@href[1]{\endgroup#1\@@endlink}%
\providecommand \@sanitize@url [0]{\catcode `\\12\catcode `\$12\catcode `\&12\catcode `\#12\catcode `\^12\catcode `\_12\catcode `\%12\relax}%
\providecommand \@@startlink[1]{}%
\providecommand \@@endlink[0]{}%
\providecommand \url  [0]{\begingroup\@sanitize@url \@url }%
\providecommand \@url [1]{\endgroup\@href {#1}{\urlprefix }}%
\providecommand \urlprefix  [0]{URL }%
\providecommand \Eprint [0]{\href }%
\providecommand \doibase [0]{http://dx.doi.org/}%
\providecommand \selectlanguage [0]{\@gobble}%
\providecommand \bibinfo  [0]{\@secondoftwo}%
\providecommand \bibfield  [0]{\@secondoftwo}%
\providecommand \translation [1]{[#1]}%
\providecommand \BibitemOpen [0]{}%
\providecommand \bibitemStop [0]{}%
\providecommand \bibitemNoStop [0]{.\EOS\space}%
\providecommand \EOS [0]{\spacefactor3000\relax}%
\providecommand \BibitemShut  [1]{\csname bibitem#1\endcsname}%
\let\auto@bib@innerbib\@empty
\bibitem [{\citenamefont {Yuan}\ \emph {et~al.}(2010)\citenamefont {Yuan}, \citenamefont {Bao}, \citenamefont {Lu}, \citenamefont {Zhang}, \citenamefont {Peng},\ and\ \citenamefont {Pan}}]{yuan_entangled_2010}%
  \BibitemOpen
  \bibfield  {author} {\bibinfo {author} {\bibfnamefont {Z.-S.}\ \bibnamefont {Yuan}}, \bibinfo {author} {\bibfnamefont {X.-H.}\ \bibnamefont {Bao}}, \bibinfo {author} {\bibfnamefont {C.-Y.}\ \bibnamefont {Lu}}, \bibinfo {author} {\bibfnamefont {J.}~\bibnamefont {Zhang}}, \bibinfo {author} {\bibfnamefont {C.-Z.}\ \bibnamefont {Peng}}, \ and\ \bibinfo {author} {\bibfnamefont {J.-W.}\ \bibnamefont {Pan}},\ }\href {\doibase 10.1016/j.physrep.2010.07.004} {\bibfield  {journal} {\bibinfo  {journal} {Physics Reports}\ }\textbf {\bibinfo {volume} {497}},\ \bibinfo {pages} {1} (\bibinfo {year} {2010})}\BibitemShut {NoStop}%
\bibitem [{\citenamefont {Flamini}\ \emph {et~al.}(2018)\citenamefont {Flamini}, \citenamefont {Spagnolo},\ and\ \citenamefont {Sciarrino}}]{flamini_photonic_2018}%
  \BibitemOpen
  \bibfield  {author} {\bibinfo {author} {\bibfnamefont {F.}~\bibnamefont {Flamini}}, \bibinfo {author} {\bibfnamefont {N.}~\bibnamefont {Spagnolo}}, \ and\ \bibinfo {author} {\bibfnamefont {F.}~\bibnamefont {Sciarrino}},\ }\href {\doibase 10.1088/1361-6633/aad5b2} {\bibfield  {journal} {\bibinfo  {journal} {Reports on Progress in Physics}\ }\textbf {\bibinfo {volume} {82}},\ \bibinfo {pages} {016001} (\bibinfo {year} {2018})},\ \bibinfo {note} {publisher: IOP Publishing}\BibitemShut {NoStop}%
\bibitem [{\citenamefont {Gühne}\ and\ \citenamefont {Tóth}(2009)}]{guhne_entanglement_2009}%
  \BibitemOpen
  \bibfield  {author} {\bibinfo {author} {\bibfnamefont {O.}~\bibnamefont {Gühne}}\ and\ \bibinfo {author} {\bibfnamefont {G.}~\bibnamefont {Tóth}},\ }\href {\doibase 10.1016/j.physrep.2009.02.004} {\bibfield  {journal} {\bibinfo  {journal} {Physics Reports}\ }\textbf {\bibinfo {volume} {474}},\ \bibinfo {pages} {1} (\bibinfo {year} {2009})}\BibitemShut {NoStop}%
\bibitem [{\citenamefont {Walter}\ \emph {et~al.}(2016)\citenamefont {Walter}, \citenamefont {Gross},\ and\ \citenamefont {Eisert}}]{walter_multipartite_2016}%
  \BibitemOpen
  \bibfield  {author} {\bibinfo {author} {\bibfnamefont {M.}~\bibnamefont {Walter}}, \bibinfo {author} {\bibfnamefont {D.}~\bibnamefont {Gross}}, \ and\ \bibinfo {author} {\bibfnamefont {J.}~\bibnamefont {Eisert}},\ }in\ \href {https://onlinelibrary.wiley.com/doi/abs/10.1002/9783527805785.ch14} {{\selectlanguage {en}\emph {\bibinfo {booktitle} {Quantum {Information}}}}}\ (\bibinfo  {publisher} {John Wiley \& Sons, Ltd},\ \bibinfo {year} {2016})\ pp.\ \bibinfo {pages} {293--330}\BibitemShut {NoStop}%
\bibitem [{\citenamefont {Gimeno-Segovia}\ \emph {et~al.}(2015)\citenamefont {Gimeno-Segovia}, \citenamefont {Shadbolt}, \citenamefont {Browne},\ and\ \citenamefont {Rudolph}}]{gimeno-segovia_three-photon_2015}%
  \BibitemOpen
  \bibfield  {author} {\bibinfo {author} {\bibfnamefont {M.}~\bibnamefont {Gimeno-Segovia}}, \bibinfo {author} {\bibfnamefont {P.}~\bibnamefont {Shadbolt}}, \bibinfo {author} {\bibfnamefont {D.~E.}\ \bibnamefont {Browne}}, \ and\ \bibinfo {author} {\bibfnamefont {T.}~\bibnamefont {Rudolph}},\ }\href {\doibase 10.1103/PhysRevLett.115.020502} {\bibfield  {journal} {\bibinfo  {journal} {Physical Review Letters}\ }\textbf {\bibinfo {volume} {115}},\ \bibinfo {pages} {020502} (\bibinfo {year} {2015})},\ \bibinfo {note} {publisher: American Physical Society}\BibitemShut {NoStop}%
\bibitem [{\citenamefont {Cleve}\ and\ \citenamefont {Buhrman}(1997)}]{cleve_substituting_1997}%
  \BibitemOpen
  \bibfield  {author} {\bibinfo {author} {\bibfnamefont {R.}~\bibnamefont {Cleve}}\ and\ \bibinfo {author} {\bibfnamefont {H.}~\bibnamefont {Buhrman}},\ }\href {\doibase 10.1103/PhysRevA.56.1201} {\bibfield  {journal} {\bibinfo  {journal} {Physical Review A}\ }\textbf {\bibinfo {volume} {56}},\ \bibinfo {pages} {1201} (\bibinfo {year} {1997})},\ \bibinfo {note} {publisher: American Physical Society}\BibitemShut {NoStop}%
\bibitem [{\citenamefont {Raussendorf}\ and\ \citenamefont {Briegel}(2001)}]{raussendorf_one-way_2001}%
  \BibitemOpen
  \bibfield  {author} {\bibinfo {author} {\bibfnamefont {R.}~\bibnamefont {Raussendorf}}\ and\ \bibinfo {author} {\bibfnamefont {H.~J.}\ \bibnamefont {Briegel}},\ }\href {\doibase 10.1103/PhysRevLett.86.5188} {\bibfield  {journal} {\bibinfo  {journal} {Physical Review Letters}\ }\textbf {\bibinfo {volume} {86}},\ \bibinfo {pages} {5188} (\bibinfo {year} {2001})}\BibitemShut {NoStop}%
\bibitem [{\citenamefont {Greenberger}\ \emph {et~al.}(1989)\citenamefont {Greenberger}, \citenamefont {Horne},\ and\ \citenamefont {Zeilinger}}]{greenberger_going_1989}%
  \BibitemOpen
  \bibfield  {author} {\bibinfo {author} {\bibfnamefont {D.~M.}\ \bibnamefont {Greenberger}}, \bibinfo {author} {\bibfnamefont {M.~A.}\ \bibnamefont {Horne}}, \ and\ \bibinfo {author} {\bibfnamefont {A.}~\bibnamefont {Zeilinger}},\ }in\ \href {\doibase 10.1007/978-94-017-0849-4_10} {{\selectlanguage {en}\emph {\bibinfo {booktitle} {Bell's {Theorem}, {Quantum} {Theory} and {Conceptions} of the {Universe}}}}},\ \bibinfo {editor} {edited by\ \bibinfo {editor} {\bibfnamefont {M.}~\bibnamefont {Kafatos}}}\ (\bibinfo  {publisher} {Springer Netherlands},\ \bibinfo {address} {Dordrecht},\ \bibinfo {year} {1989})\ pp.\ \bibinfo {pages} {69--72}\BibitemShut {NoStop}%
\bibitem [{\citenamefont {Dür}\ \emph {et~al.}(2000)\citenamefont {Dür}, \citenamefont {Vidal},\ and\ \citenamefont {Cirac}}]{dur_three_2000}%
  \BibitemOpen
  \bibfield  {author} {\bibinfo {author} {\bibfnamefont {W.}~\bibnamefont {Dür}}, \bibinfo {author} {\bibfnamefont {G.}~\bibnamefont {Vidal}}, \ and\ \bibinfo {author} {\bibfnamefont {J.~I.}\ \bibnamefont {Cirac}},\ }\href {\doibase 10.1103/PhysRevA.62.062314} {\bibfield  {journal} {\bibinfo  {journal} {Physical Review A}\ }\textbf {\bibinfo {volume} {62}},\ \bibinfo {pages} {062314} (\bibinfo {year} {2000})},\ \bibinfo {note} {publisher: American Physical Society}\BibitemShut {NoStop}%
\bibitem [{\citenamefont {Joo}\ \emph {et~al.}(2003)\citenamefont {Joo}, \citenamefont {Park}, \citenamefont {Oh},\ and\ \citenamefont {Kim}}]{joo_quantum_2003}%
  \BibitemOpen
  \bibfield  {author} {\bibinfo {author} {\bibfnamefont {J.}~\bibnamefont {Joo}}, \bibinfo {author} {\bibfnamefont {Y.-J.}\ \bibnamefont {Park}}, \bibinfo {author} {\bibfnamefont {S.}~\bibnamefont {Oh}}, \ and\ \bibinfo {author} {\bibfnamefont {J.}~\bibnamefont {Kim}},\ }\href {\doibase 10.1088/1367-2630/5/1/136} {\bibfield  {journal} {\bibinfo  {journal} {New Journal of Physics}\ }\textbf {\bibinfo {volume} {5}},\ \bibinfo {pages} {136} (\bibinfo {year} {2003})}\BibitemShut {NoStop}%
\bibitem [{\citenamefont {Jian}\ \emph {et~al.}(2007)\citenamefont {Jian}, \citenamefont {Quan},\ and\ \citenamefont {Chao-Jing}}]{jian_quantum_2007}%
  \BibitemOpen
  \bibfield  {author} {\bibinfo {author} {\bibfnamefont {W.}~\bibnamefont {Jian}}, \bibinfo {author} {\bibfnamefont {Z.}~\bibnamefont {Quan}}, \ and\ \bibinfo {author} {\bibfnamefont {T.}~\bibnamefont {Chao-Jing}},\ }\href {\doibase 10.1088/0253-6102/48/4/013} {\bibfield  {journal} {\bibinfo  {journal} {Communications in Theoretical Physics}\ }\textbf {\bibinfo {volume} {48}},\ \bibinfo {pages} {637} (\bibinfo {year} {2007})}\BibitemShut {NoStop}%
\bibitem [{\citenamefont {Singh}\ \emph {et~al.}(2016)\citenamefont {Singh}, \citenamefont {Adhikari},\ and\ \citenamefont {Kumar}}]{singh_usefulness_2016}%
  \BibitemOpen
  \bibfield  {author} {\bibinfo {author} {\bibfnamefont {P.}~\bibnamefont {Singh}}, \bibinfo {author} {\bibfnamefont {S.}~\bibnamefont {Adhikari}}, \ and\ \bibinfo {author} {\bibfnamefont {A.}~\bibnamefont {Kumar}},\ }\href {\doibase 10.1134/S1063776116110200} {\bibfield  {journal} {\bibinfo  {journal} {Journal of Experimental and Theoretical Physics}\ }\textbf {\bibinfo {volume} {123}},\ \bibinfo {pages} {572} (\bibinfo {year} {2016})}\BibitemShut {NoStop}%
\bibitem [{\citenamefont {Cabello}(2002)}]{cabello_bells_2002}%
  \BibitemOpen
  \bibfield  {author} {\bibinfo {author} {\bibfnamefont {A.}~\bibnamefont {Cabello}},\ }\href {\doibase 10.1103/PhysRevA.65.032108} {\bibfield  {journal} {\bibinfo  {journal} {Physical Review A}\ }\textbf {\bibinfo {volume} {65}},\ \bibinfo {pages} {032108} (\bibinfo {year} {2002})},\ \bibinfo {note} {publisher: American Physical Society}\BibitemShut {NoStop}%
\bibitem [{\citenamefont {Gottesman}\ \emph {et~al.}(2012)\citenamefont {Gottesman}, \citenamefont {Jennewein},\ and\ \citenamefont {Croke}}]{gottesman_longer-baseline_2012}%
  \BibitemOpen
  \bibfield  {author} {\bibinfo {author} {\bibfnamefont {D.}~\bibnamefont {Gottesman}}, \bibinfo {author} {\bibfnamefont {T.}~\bibnamefont {Jennewein}}, \ and\ \bibinfo {author} {\bibfnamefont {S.}~\bibnamefont {Croke}},\ }\href {\doibase 10.1103/PhysRevLett.109.070503} {\bibfield  {journal} {\bibinfo  {journal} {Physical Review Letters}\ }\textbf {\bibinfo {volume} {109}},\ \bibinfo {pages} {070503} (\bibinfo {year} {2012})},\ \bibinfo {note} {publisher: American Physical Society}\BibitemShut {NoStop}%
\bibitem [{\citenamefont {Agrawal}\ and\ \citenamefont {Pati}(2006)}]{agrawal_perfect_2006}%
  \BibitemOpen
  \bibfield  {author} {\bibinfo {author} {\bibfnamefont {P.}~\bibnamefont {Agrawal}}\ and\ \bibinfo {author} {\bibfnamefont {A.}~\bibnamefont {Pati}},\ }\href {\doibase 10.1103/PhysRevA.74.062320} {\bibfield  {journal} {\bibinfo  {journal} {Physical Review A}\ }\textbf {\bibinfo {volume} {74}},\ \bibinfo {pages} {062320} (\bibinfo {year} {2006})},\ \bibinfo {note} {publisher: American Physical Society}\BibitemShut {NoStop}%
\bibitem [{\citenamefont {Yamamoto}\ \emph {et~al.}(2002)\citenamefont {Yamamoto}, \citenamefont {Tamaki}, \citenamefont {Koashi},\ and\ \citenamefont {Imoto}}]{yamamoto_polarization-entangled_2002}%
  \BibitemOpen
  \bibfield  {author} {\bibinfo {author} {\bibfnamefont {T.}~\bibnamefont {Yamamoto}}, \bibinfo {author} {\bibfnamefont {K.}~\bibnamefont {Tamaki}}, \bibinfo {author} {\bibfnamefont {M.}~\bibnamefont {Koashi}}, \ and\ \bibinfo {author} {\bibfnamefont {N.}~\bibnamefont {Imoto}},\ }\href {\doibase 10.1103/PhysRevA.66.064301} {\bibfield  {journal} {\bibinfo  {journal} {Physical Review A}\ }\textbf {\bibinfo {volume} {66}},\ \bibinfo {pages} {064301} (\bibinfo {year} {2002})},\ \bibinfo {note} {publisher: American Physical Society}\BibitemShut {NoStop}%
\bibitem [{\citenamefont {Eibl}\ \emph {et~al.}(2004)\citenamefont {Eibl}, \citenamefont {Kiesel}, \citenamefont {Bourennane}, \citenamefont {Kurtsiefer},\ and\ \citenamefont {Weinfurter}}]{eibl_experimental_2004}%
  \BibitemOpen
  \bibfield  {author} {\bibinfo {author} {\bibfnamefont {M.}~\bibnamefont {Eibl}}, \bibinfo {author} {\bibfnamefont {N.}~\bibnamefont {Kiesel}}, \bibinfo {author} {\bibfnamefont {M.}~\bibnamefont {Bourennane}}, \bibinfo {author} {\bibfnamefont {C.}~\bibnamefont {Kurtsiefer}}, \ and\ \bibinfo {author} {\bibfnamefont {H.}~\bibnamefont {Weinfurter}},\ }\href {\doibase 10.1103/PhysRevLett.92.077901} {\bibfield  {journal} {\bibinfo  {journal} {Physical Review Letters}\ }\textbf {\bibinfo {volume} {92}},\ \bibinfo {pages} {077901} (\bibinfo {year} {2004})}\BibitemShut {NoStop}%
\bibitem [{\citenamefont {Gräfe}\ \emph {et~al.}(2014)\citenamefont {Gräfe}, \citenamefont {Heilmann}, \citenamefont {Perez-Leija}, \citenamefont {Keil}, \citenamefont {Dreisow}, \citenamefont {Heinrich}, \citenamefont {Moya-Cessa}, \citenamefont {Nolte}, \citenamefont {Christodoulides},\ and\ \citenamefont {Szameit}}]{grafe_-chip_2014}%
  \BibitemOpen
  \bibfield  {author} {\bibinfo {author} {\bibfnamefont {M.}~\bibnamefont {Gräfe}}, \bibinfo {author} {\bibfnamefont {R.}~\bibnamefont {Heilmann}}, \bibinfo {author} {\bibfnamefont {A.}~\bibnamefont {Perez-Leija}}, \bibinfo {author} {\bibfnamefont {R.}~\bibnamefont {Keil}}, \bibinfo {author} {\bibfnamefont {F.}~\bibnamefont {Dreisow}}, \bibinfo {author} {\bibfnamefont {M.}~\bibnamefont {Heinrich}}, \bibinfo {author} {\bibfnamefont {H.}~\bibnamefont {Moya-Cessa}}, \bibinfo {author} {\bibfnamefont {S.}~\bibnamefont {Nolte}}, \bibinfo {author} {\bibfnamefont {D.~N.}\ \bibnamefont {Christodoulides}}, \ and\ \bibinfo {author} {\bibfnamefont {A.}~\bibnamefont {Szameit}},\ }\href {\doibase 10.1038/nphoton.2014.204} {\bibfield  {journal} {\bibinfo  {journal} {Nature Photonics}\ }\textbf {\bibinfo {volume} {8}},\ \bibinfo {pages} {791} (\bibinfo {year} {2014})},\ \bibinfo {note} {number: 10 Publisher: Nature Publishing Group}\BibitemShut {NoStop}%
\bibitem [{\citenamefont {Zheng}\ \emph {et~al.}(2022)\citenamefont {Zheng}, \citenamefont {Peng}, \citenamefont {Tang}, \citenamefont {Li},\ and\ \citenamefont {Tan}}]{zheng_unified_2022}%
  \BibitemOpen
  \bibfield  {author} {\bibinfo {author} {\bibfnamefont {J.}~\bibnamefont {Zheng}}, \bibinfo {author} {\bibfnamefont {J.}~\bibnamefont {Peng}}, \bibinfo {author} {\bibfnamefont {P.}~\bibnamefont {Tang}}, \bibinfo {author} {\bibfnamefont {F.}~\bibnamefont {Li}}, \ and\ \bibinfo {author} {\bibfnamefont {N.}~\bibnamefont {Tan}},\ }\href {\doibase 10.1103/PhysRevA.105.062408} {\bibfield  {journal} {\bibinfo  {journal} {Physical Review A}\ }\textbf {\bibinfo {volume} {105}},\ \bibinfo {pages} {062408} (\bibinfo {year} {2022})},\ \bibinfo {note} {publisher: American Physical Society}\BibitemShut {NoStop}%
\bibitem [{\citenamefont {Swain}\ \emph {et~al.}(2023)\citenamefont {Swain}, \citenamefont {Selvan}, \citenamefont {Rai},\ and\ \citenamefont {Panigrahi}}]{swain_generation_2023}%
  \BibitemOpen
  \bibfield  {author} {\bibinfo {author} {\bibfnamefont {M.}~\bibnamefont {Swain}}, \bibinfo {author} {\bibfnamefont {M.~K.}\ \bibnamefont {Selvan}}, \bibinfo {author} {\bibfnamefont {A.}~\bibnamefont {Rai}}, \ and\ \bibinfo {author} {\bibfnamefont {P.~K.}\ \bibnamefont {Panigrahi}},\ }\href {\doibase 10.1007/s11128-023-04057-3} {\bibfield  {journal} {\bibinfo  {journal} {Quantum Information Processing}\ }\textbf {\bibinfo {volume} {22}},\ \bibinfo {pages} {302} (\bibinfo {year} {2023})}\BibitemShut {NoStop}%
\bibitem [{\citenamefont {Bao}\ \emph {et~al.}(2023)\citenamefont {Bao}, \citenamefont {Fu}, \citenamefont {Pramanik}, \citenamefont {Mao}, \citenamefont {Chi}, \citenamefont {Cao}, \citenamefont {Zhai}, \citenamefont {Mao}, \citenamefont {Dai}, \citenamefont {Chen}, \citenamefont {Jia}, \citenamefont {Zhao}, \citenamefont {Zheng}, \citenamefont {Tang}, \citenamefont {Li}, \citenamefont {Luo}, \citenamefont {Wang}, \citenamefont {Yang}, \citenamefont {Peng}, \citenamefont {Liu}, \citenamefont {Dai}, \citenamefont {He}, \citenamefont {Muthali}, \citenamefont {Oxenløwe}, \citenamefont {Vigliar}, \citenamefont {Paesani}, \citenamefont {Hou}, \citenamefont {Santagati}, \citenamefont {Silverstone}, \citenamefont {Laing}, \citenamefont {Thompson}, \citenamefont {O'Brien}, \citenamefont {Ding}, \citenamefont {Gong},\ and\ \citenamefont {Wang}}]{bao_very-large-scale_2023}%
  \BibitemOpen
  \bibfield  {author} {\bibinfo {author} {\bibfnamefont {J.}~\bibnamefont {Bao}}, \bibinfo {author} {\bibfnamefont {Z.}~\bibnamefont {Fu}}, \bibinfo {author} {\bibfnamefont {T.}~\bibnamefont {Pramanik}}, \bibinfo {author} {\bibfnamefont {J.}~\bibnamefont {Mao}}, \bibinfo {author} {\bibfnamefont {Y.}~\bibnamefont {Chi}}, \bibinfo {author} {\bibfnamefont {Y.}~\bibnamefont {Cao}}, \bibinfo {author} {\bibfnamefont {C.}~\bibnamefont {Zhai}}, \bibinfo {author} {\bibfnamefont {Y.}~\bibnamefont {Mao}}, \bibinfo {author} {\bibfnamefont {T.}~\bibnamefont {Dai}}, \bibinfo {author} {\bibfnamefont {X.}~\bibnamefont {Chen}}, \bibinfo {author} {\bibfnamefont {X.}~\bibnamefont {Jia}}, \bibinfo {author} {\bibfnamefont {L.}~\bibnamefont {Zhao}}, \bibinfo {author} {\bibfnamefont {Y.}~\bibnamefont {Zheng}}, \bibinfo {author} {\bibfnamefont {B.}~\bibnamefont {Tang}}, \bibinfo {author} {\bibfnamefont {Z.}~\bibnamefont {Li}}, \bibinfo {author} {\bibfnamefont {J.}~\bibnamefont {Luo}}, \bibinfo {author} {\bibfnamefont
  {W.}~\bibnamefont {Wang}}, \bibinfo {author} {\bibfnamefont {Y.}~\bibnamefont {Yang}}, \bibinfo {author} {\bibfnamefont {Y.}~\bibnamefont {Peng}}, \bibinfo {author} {\bibfnamefont {D.}~\bibnamefont {Liu}}, \bibinfo {author} {\bibfnamefont {D.}~\bibnamefont {Dai}}, \bibinfo {author} {\bibfnamefont {Q.}~\bibnamefont {He}}, \bibinfo {author} {\bibfnamefont {A.~L.}\ \bibnamefont {Muthali}}, \bibinfo {author} {\bibfnamefont {L.~K.}\ \bibnamefont {Oxenløwe}}, \bibinfo {author} {\bibfnamefont {C.}~\bibnamefont {Vigliar}}, \bibinfo {author} {\bibfnamefont {S.}~\bibnamefont {Paesani}}, \bibinfo {author} {\bibfnamefont {H.}~\bibnamefont {Hou}}, \bibinfo {author} {\bibfnamefont {R.}~\bibnamefont {Santagati}}, \bibinfo {author} {\bibfnamefont {J.~W.}\ \bibnamefont {Silverstone}}, \bibinfo {author} {\bibfnamefont {A.}~\bibnamefont {Laing}}, \bibinfo {author} {\bibfnamefont {M.~G.}\ \bibnamefont {Thompson}}, \bibinfo {author} {\bibfnamefont {J.~L.}\ \bibnamefont {O'Brien}}, \bibinfo {author} {\bibfnamefont
  {Y.}~\bibnamefont {Ding}}, \bibinfo {author} {\bibfnamefont {Q.}~\bibnamefont {Gong}}, \ and\ \bibinfo {author} {\bibfnamefont {J.}~\bibnamefont {Wang}},\ }\href {\doibase 10.1038/s41566-023-01187-z} {\bibfield  {journal} {\bibinfo  {journal} {Nature Photonics}\ }\textbf {\bibinfo {volume} {17}},\ \bibinfo {pages} {573} (\bibinfo {year} {2023})},\ \bibinfo {note} {publisher: Nature Publishing Group}\BibitemShut {NoStop}%
\bibitem [{\citenamefont {Besse}\ \emph {et~al.}(2020)\citenamefont {Besse}, \citenamefont {Reuer}, \citenamefont {Collodo}, \citenamefont {Wulff}, \citenamefont {Wernli}, \citenamefont {Copetudo}, \citenamefont {Malz}, \citenamefont {Magnard}, \citenamefont {Akin}, \citenamefont {Gabureac}, \citenamefont {Norris}, \citenamefont {Cirac}, \citenamefont {Wallraff},\ and\ \citenamefont {Eichler}}]{besse_realizing_2020}%
  \BibitemOpen
  \bibfield  {author} {\bibinfo {author} {\bibfnamefont {J.-C.}\ \bibnamefont {Besse}}, \bibinfo {author} {\bibfnamefont {K.}~\bibnamefont {Reuer}}, \bibinfo {author} {\bibfnamefont {M.~C.}\ \bibnamefont {Collodo}}, \bibinfo {author} {\bibfnamefont {A.}~\bibnamefont {Wulff}}, \bibinfo {author} {\bibfnamefont {L.}~\bibnamefont {Wernli}}, \bibinfo {author} {\bibfnamefont {A.}~\bibnamefont {Copetudo}}, \bibinfo {author} {\bibfnamefont {D.}~\bibnamefont {Malz}}, \bibinfo {author} {\bibfnamefont {P.}~\bibnamefont {Magnard}}, \bibinfo {author} {\bibfnamefont {A.}~\bibnamefont {Akin}}, \bibinfo {author} {\bibfnamefont {M.}~\bibnamefont {Gabureac}}, \bibinfo {author} {\bibfnamefont {G.~J.}\ \bibnamefont {Norris}}, \bibinfo {author} {\bibfnamefont {J.~I.}\ \bibnamefont {Cirac}}, \bibinfo {author} {\bibfnamefont {A.}~\bibnamefont {Wallraff}}, \ and\ \bibinfo {author} {\bibfnamefont {C.}~\bibnamefont {Eichler}},\ }\href {\doibase 10.1038/s41467-020-18635-x} {\bibfield  {journal} {\bibinfo  {journal} {Nature
  Communications}\ }\textbf {\bibinfo {volume} {11}},\ \bibinfo {pages} {4877} (\bibinfo {year} {2020})}\BibitemShut {NoStop}%
\bibitem [{\citenamefont {Xiao}\ \emph {et~al.}(2008)\citenamefont {Xiao}, \citenamefont {Wang}, \citenamefont {Zhang}, \citenamefont {Huang}, \citenamefont {Peng},\ and\ \citenamefont {Long}}]{xiao_efficient_2008}%
  \BibitemOpen
  \bibfield  {author} {\bibinfo {author} {\bibfnamefont {L.}~\bibnamefont {Xiao}}, \bibinfo {author} {\bibfnamefont {C.}~\bibnamefont {Wang}}, \bibinfo {author} {\bibfnamefont {W.}~\bibnamefont {Zhang}}, \bibinfo {author} {\bibfnamefont {Y.}~\bibnamefont {Huang}}, \bibinfo {author} {\bibfnamefont {J.}~\bibnamefont {Peng}}, \ and\ \bibinfo {author} {\bibfnamefont {G.}~\bibnamefont {Long}},\ }\href {\doibase 10.1103/PhysRevA.77.042315} {\bibfield  {journal} {\bibinfo  {journal} {Physical Review A}\ }\textbf {\bibinfo {volume} {77}},\ \bibinfo {pages} {042315} (\bibinfo {year} {2008})},\ \bibinfo {note} {publisher: American Physical Society}\BibitemShut {NoStop}%
\bibitem [{\citenamefont {Antonelli}\ \emph {et~al.}(2011)\citenamefont {Antonelli}, \citenamefont {Shtaif},\ and\ \citenamefont {Brodsky}}]{antonelli_sudden_2011}%
  \BibitemOpen
  \bibfield  {author} {\bibinfo {author} {\bibfnamefont {C.}~\bibnamefont {Antonelli}}, \bibinfo {author} {\bibfnamefont {M.}~\bibnamefont {Shtaif}}, \ and\ \bibinfo {author} {\bibfnamefont {M.}~\bibnamefont {Brodsky}},\ }\href {\doibase 10.1103/PhysRevLett.106.080404} {\bibfield  {journal} {\bibinfo  {journal} {Physical Review Letters}\ }\textbf {\bibinfo {volume} {106}},\ \bibinfo {pages} {080404} (\bibinfo {year} {2011})},\ \bibinfo {note} {arXiv:1101.5417 [quant-ph]}\BibitemShut {NoStop}%
\bibitem [{\citenamefont {Menotti}\ \emph {et~al.}(2016)\citenamefont {Menotti}, \citenamefont {Maccone}, \citenamefont {Sipe},\ and\ \citenamefont {Liscidini}}]{menotti_generation_2016}%
  \BibitemOpen
  \bibfield  {author} {\bibinfo {author} {\bibfnamefont {M.}~\bibnamefont {Menotti}}, \bibinfo {author} {\bibfnamefont {L.}~\bibnamefont {Maccone}}, \bibinfo {author} {\bibfnamefont {J.~E.}\ \bibnamefont {Sipe}}, \ and\ \bibinfo {author} {\bibfnamefont {M.}~\bibnamefont {Liscidini}},\ }\href {\doibase 10.1103/PhysRevA.94.013845} {\bibfield  {journal} {\bibinfo  {journal} {Physical Review A}\ }\textbf {\bibinfo {volume} {94}},\ \bibinfo {pages} {013845} (\bibinfo {year} {2016})},\ \bibinfo {note} {publisher: American Physical Society}\BibitemShut {NoStop}%
\bibitem [{\citenamefont {Fang}\ \emph {et~al.}(2019)\citenamefont {Fang}, \citenamefont {Menotti}, \citenamefont {Liscidini}, \citenamefont {Sipe},\ and\ \citenamefont {Lorenz}}]{fang_three-photon_2019}%
  \BibitemOpen
  \bibfield  {author} {\bibinfo {author} {\bibfnamefont {B.}~\bibnamefont {Fang}}, \bibinfo {author} {\bibfnamefont {M.}~\bibnamefont {Menotti}}, \bibinfo {author} {\bibfnamefont {M.}~\bibnamefont {Liscidini}}, \bibinfo {author} {\bibfnamefont {J.}~\bibnamefont {Sipe}}, \ and\ \bibinfo {author} {\bibfnamefont {V.}~\bibnamefont {Lorenz}},\ }\href {\doibase 10.1103/PhysRevLett.123.070508} {\bibfield  {journal} {\bibinfo  {journal} {Physical Review Letters}\ }\textbf {\bibinfo {volume} {123}},\ \bibinfo {pages} {070508} (\bibinfo {year} {2019})},\ \bibinfo {note} {publisher: American Physical Society}\BibitemShut {NoStop}%
\bibitem [{\citenamefont {Banic}\ \emph {et~al.}(2024)\citenamefont {Banic}, \citenamefont {Sipe},\ and\ \citenamefont {Liscidini}}]{banic_integrated_2024}%
  \BibitemOpen
  \bibfield  {author} {\bibinfo {author} {\bibfnamefont {M.}~\bibnamefont {Banic}}, \bibinfo {author} {\bibfnamefont {J.~E.}\ \bibnamefont {Sipe}}, \ and\ \bibinfo {author} {\bibfnamefont {M.}~\bibnamefont {Liscidini}},\ }\href {\doibase 10.1103/PhysRevA.109.013505} {\bibfield  {journal} {\bibinfo  {journal} {Physical Review A}\ }\textbf {\bibinfo {volume} {109}},\ \bibinfo {pages} {013505} (\bibinfo {year} {2024})},\ \bibinfo {note} {publisher: American Physical Society}\BibitemShut {NoStop}%
\bibitem [{\citenamefont {McGuinness}\ \emph {et~al.}(2011)\citenamefont {McGuinness}, \citenamefont {Raymer},\ and\ \citenamefont {McKinstrie}}]{mcguinness_theory_2011}%
  \BibitemOpen
  \bibfield  {author} {\bibinfo {author} {\bibfnamefont {H.~J.}\ \bibnamefont {McGuinness}}, \bibinfo {author} {\bibfnamefont {M.~G.}\ \bibnamefont {Raymer}}, \ and\ \bibinfo {author} {\bibfnamefont {C.~J.}\ \bibnamefont {McKinstrie}},\ }\href {\doibase 10.1364/OE.19.017876} {\bibfield  {journal} {\bibinfo  {journal} {Optics Express}\ }\textbf {\bibinfo {volume} {19}},\ \bibinfo {pages} {17876} (\bibinfo {year} {2011})},\ \bibinfo {note} {publisher: Optica Publishing Group}\BibitemShut {NoStop}%
\bibitem [{\citenamefont {Behunin}\ and\ \citenamefont {Rakich}(2023)}]{behunin_harnessing_2023}%
  \BibitemOpen
  \bibfield  {author} {\bibinfo {author} {\bibfnamefont {R.~O.}\ \bibnamefont {Behunin}}\ and\ \bibinfo {author} {\bibfnamefont {P.~T.}\ \bibnamefont {Rakich}},\ }\href {\doibase 10.1103/PhysRevA.107.023511} {\bibfield  {journal} {\bibinfo  {journal} {Physical Review A}\ }\textbf {\bibinfo {volume} {107}},\ \bibinfo {pages} {023511} (\bibinfo {year} {2023})},\ \bibinfo {note} {publisher: American Physical Society}\BibitemShut {NoStop}%
\bibitem [{\citenamefont {Kharel}\ \emph {et~al.}(2016)\citenamefont {Kharel}, \citenamefont {Behunin}, \citenamefont {Renninger},\ and\ \citenamefont {Rakich}}]{kharel_noise_2016}%
  \BibitemOpen
  \bibfield  {author} {\bibinfo {author} {\bibfnamefont {P.}~\bibnamefont {Kharel}}, \bibinfo {author} {\bibfnamefont {R.~O.}\ \bibnamefont {Behunin}}, \bibinfo {author} {\bibfnamefont {W.~H.}\ \bibnamefont {Renninger}}, \ and\ \bibinfo {author} {\bibfnamefont {P.~T.}\ \bibnamefont {Rakich}},\ }\href {\doibase 10.1103/PhysRevA.93.063806} {\bibfield  {journal} {\bibinfo  {journal} {Physical Review A}\ }\textbf {\bibinfo {volume} {93}},\ \bibinfo {pages} {063806} (\bibinfo {year} {2016})},\ \bibinfo {note} {publisher: American Physical Society}\BibitemShut {NoStop}%
\bibitem [{\citenamefont {Shepherd}\ and\ \citenamefont {Behunin}(2024)}]{shepherd_multi-phonon_2024}%
  \BibitemOpen
  \bibfield  {author} {\bibinfo {author} {\bibfnamefont {A.~J.}\ \bibnamefont {Shepherd}}\ and\ \bibinfo {author} {\bibfnamefont {R.~O.}\ \bibnamefont {Behunin}},\ }\href {http://arxiv.org/abs/2407.19120} {{\selectlanguage {en}\enquote {\bibinfo {title} {Multi-phonon {Fock} state heralding with single-photon detection},}\ }} (\bibinfo {year} {2024}),\ \bibinfo {note} {arXiv:2407.19120 [physics, physics:quant-ph]}\BibitemShut {NoStop}%
\bibitem [{\citenamefont {Liu}\ \emph {et~al.}(2024{\natexlab{a}})\citenamefont {Liu}, \citenamefont {Wang}, \citenamefont {Chauhan}, \citenamefont {Harrington}, \citenamefont {Nelson},\ and\ \citenamefont {Blumenthal}}]{liu_integrated_2024}%
  \BibitemOpen
  \bibfield  {author} {\bibinfo {author} {\bibfnamefont {K.}~\bibnamefont {Liu}}, \bibinfo {author} {\bibfnamefont {J.}~\bibnamefont {Wang}}, \bibinfo {author} {\bibfnamefont {N.}~\bibnamefont {Chauhan}}, \bibinfo {author} {\bibfnamefont {M.~W.}\ \bibnamefont {Harrington}}, \bibinfo {author} {\bibfnamefont {K.~D.}\ \bibnamefont {Nelson}}, \ and\ \bibinfo {author} {\bibfnamefont {D.~J.}\ \bibnamefont {Blumenthal}},\ }\href {\doibase 10.1364/OL.503126} {\bibfield  {journal} {\bibinfo  {journal} {Optics Letters}\ }\textbf {\bibinfo {volume} {49}},\ \bibinfo {pages} {45} (\bibinfo {year} {2024}{\natexlab{a}})},\ \bibinfo {note} {publisher: Optica Publishing Group}\BibitemShut {NoStop}%
\bibitem [{\citenamefont {Liu}\ \emph {et~al.}(2024{\natexlab{b}})\citenamefont {Liu}, \citenamefont {Chauhan}, \citenamefont {Song}, \citenamefont {Harrington}, \citenamefont {Nelson},\ and\ \citenamefont {Blumenthal}}]{liu_tunable_2024}%
  \BibitemOpen
  \bibfield  {author} {\bibinfo {author} {\bibfnamefont {K.}~\bibnamefont {Liu}}, \bibinfo {author} {\bibfnamefont {N.}~\bibnamefont {Chauhan}}, \bibinfo {author} {\bibfnamefont {M.}~\bibnamefont {Song}}, \bibinfo {author} {\bibfnamefont {M.~W.}\ \bibnamefont {Harrington}}, \bibinfo {author} {\bibfnamefont {K.~D.}\ \bibnamefont {Nelson}}, \ and\ \bibinfo {author} {\bibfnamefont {D.~J.}\ \bibnamefont {Blumenthal}},\ }\href {\doibase 10.1364/PRJ.528398} {\bibfield  {journal} {\bibinfo  {journal} {Photonics Research}\ }\textbf {\bibinfo {volume} {12}},\ \bibinfo {pages} {1890} (\bibinfo {year} {2024}{\natexlab{b}})},\ \bibinfo {note} {publisher: Optica Publishing Group}\BibitemShut {NoStop}%
\bibitem [{\citenamefont {McGuinness}\ \emph {et~al.}(2010)\citenamefont {McGuinness}, \citenamefont {Raymer}, \citenamefont {McKinstrie},\ and\ \citenamefont {Radic}}]{mcguinness_quantum_2010}%
  \BibitemOpen
  \bibfield  {author} {\bibinfo {author} {\bibfnamefont {H.~J.}\ \bibnamefont {McGuinness}}, \bibinfo {author} {\bibfnamefont {M.~G.}\ \bibnamefont {Raymer}}, \bibinfo {author} {\bibfnamefont {C.~J.}\ \bibnamefont {McKinstrie}}, \ and\ \bibinfo {author} {\bibfnamefont {S.}~\bibnamefont {Radic}},\ }\href {\doibase 10.1103/PhysRevLett.105.093604} {\bibfield  {journal} {\bibinfo  {journal} {Physical Review Letters}\ }\textbf {\bibinfo {volume} {105}},\ \bibinfo {pages} {093604} (\bibinfo {year} {2010})},\ \bibinfo {note} {publisher: American Physical Society}\BibitemShut {NoStop}%
\bibitem [{\citenamefont {Goryachev}\ \emph {et~al.}(2012)\citenamefont {Goryachev}, \citenamefont {Creedon}, \citenamefont {Ivanov}, \citenamefont {Galliou}, \citenamefont {Bourquin},\ and\ \citenamefont {Tobar}}]{goryachev_extremely_2012}%
  \BibitemOpen
  \bibfield  {author} {\bibinfo {author} {\bibfnamefont {M.}~\bibnamefont {Goryachev}}, \bibinfo {author} {\bibfnamefont {D.~L.}\ \bibnamefont {Creedon}}, \bibinfo {author} {\bibfnamefont {E.~N.}\ \bibnamefont {Ivanov}}, \bibinfo {author} {\bibfnamefont {S.}~\bibnamefont {Galliou}}, \bibinfo {author} {\bibfnamefont {R.}~\bibnamefont {Bourquin}}, \ and\ \bibinfo {author} {\bibfnamefont {M.~E.}\ \bibnamefont {Tobar}},\ }\href {\doibase 10.1063/1.4729292} {\bibfield  {journal} {\bibinfo  {journal} {Applied Physics Letters}\ }\textbf {\bibinfo {volume} {100}},\ \bibinfo {pages} {243504} (\bibinfo {year} {2012})},\ \bibinfo {note} {publisher: American Institute of Physics}\BibitemShut {NoStop}%
\bibitem [{\citenamefont {MacCabe}\ \emph {et~al.}(2020)\citenamefont {MacCabe}, \citenamefont {Ren}, \citenamefont {Luo}, \citenamefont {Cohen}, \citenamefont {Zhou}, \citenamefont {Sipahigil}, \citenamefont {Mirhosseini},\ and\ \citenamefont {Painter}}]{maccabe_nano-acoustic_2020}%
  \BibitemOpen
  \bibfield  {author} {\bibinfo {author} {\bibfnamefont {G.~S.}\ \bibnamefont {MacCabe}}, \bibinfo {author} {\bibfnamefont {H.}~\bibnamefont {Ren}}, \bibinfo {author} {\bibfnamefont {J.}~\bibnamefont {Luo}}, \bibinfo {author} {\bibfnamefont {J.~D.}\ \bibnamefont {Cohen}}, \bibinfo {author} {\bibfnamefont {H.}~\bibnamefont {Zhou}}, \bibinfo {author} {\bibfnamefont {A.}~\bibnamefont {Sipahigil}}, \bibinfo {author} {\bibfnamefont {M.}~\bibnamefont {Mirhosseini}}, \ and\ \bibinfo {author} {\bibfnamefont {O.}~\bibnamefont {Painter}},\ }\href {\doibase 10.1126/science.abc7312} {\bibfield  {journal} {\bibinfo  {journal} {Science}\ }\textbf {\bibinfo {volume} {370}},\ \bibinfo {pages} {840} (\bibinfo {year} {2020})}\BibitemShut {NoStop}%
\bibitem [{\citenamefont {Boyd}(2008)}]{boyd_chapter_2008}%
  \BibitemOpen
  \bibfield  {author} {\bibinfo {author} {\bibfnamefont {R.~W.}\ \bibnamefont {Boyd}},\ }in\ \href {\doibase 10.1016/B978-0-12-369470-6.00009-5} {\emph {\bibinfo {booktitle} {Nonlinear {Optics} ({Third} {Edition})}}},\ \bibinfo {editor} {edited by\ \bibinfo {editor} {\bibfnamefont {R.~W.}\ \bibnamefont {Boyd}}}\ (\bibinfo  {publisher} {Academic Press},\ \bibinfo {address} {Burlington},\ \bibinfo {year} {2008})\ pp.\ \bibinfo {pages} {429--471}\BibitemShut {NoStop}%
\bibitem [{\citenamefont {Shelby}\ \emph {et~al.}(1985)\citenamefont {Shelby}, \citenamefont {Levenson},\ and\ \citenamefont {Bayer}}]{shelby_resolved_1985}%
  \BibitemOpen
  \bibfield  {author} {\bibinfo {author} {\bibfnamefont {R.~M.}\ \bibnamefont {Shelby}}, \bibinfo {author} {\bibfnamefont {M.~D.}\ \bibnamefont {Levenson}}, \ and\ \bibinfo {author} {\bibfnamefont {P.~W.}\ \bibnamefont {Bayer}},\ }\href {\doibase 10.1103/PhysRevLett.54.939} {\bibfield  {journal} {\bibinfo  {journal} {Physical Review Letters}\ }\textbf {\bibinfo {volume} {54}},\ \bibinfo {pages} {939} (\bibinfo {year} {1985})},\ \bibinfo {note} {publisher: American Physical Society}\BibitemShut {NoStop}%
\bibitem [{\citenamefont {Eggleton}\ \emph {et~al.}(2019)\citenamefont {Eggleton}, \citenamefont {Poulton}, \citenamefont {Rakich}, \citenamefont {Steel},\ and\ \citenamefont {Bahl}}]{eggleton_brillouin_2019}%
  \BibitemOpen
  \bibfield  {author} {\bibinfo {author} {\bibfnamefont {B.~J.}\ \bibnamefont {Eggleton}}, \bibinfo {author} {\bibfnamefont {C.~G.}\ \bibnamefont {Poulton}}, \bibinfo {author} {\bibfnamefont {P.~T.}\ \bibnamefont {Rakich}}, \bibinfo {author} {\bibfnamefont {M.~J.}\ \bibnamefont {Steel}}, \ and\ \bibinfo {author} {\bibfnamefont {G.}~\bibnamefont {Bahl}},\ }\href {\doibase 10.1038/s41566-019-0498-z} {\bibfield  {journal} {\bibinfo  {journal} {Nature Photonics}\ }\textbf {\bibinfo {volume} {13}},\ \bibinfo {pages} {664} (\bibinfo {year} {2019})},\ \bibinfo {note} {publisher: Nature Publishing Group}\BibitemShut {NoStop}%
\bibitem [{\citenamefont {Behunin}\ \emph {et~al.}(2019)\citenamefont {Behunin}, \citenamefont {Ou},\ and\ \citenamefont {Kieu}}]{behunin_spontaneous_2019}%
  \BibitemOpen
  \bibfield  {author} {\bibinfo {author} {\bibfnamefont {R.~O.}\ \bibnamefont {Behunin}}, \bibinfo {author} {\bibfnamefont {Y.-H.}\ \bibnamefont {Ou}}, \ and\ \bibinfo {author} {\bibfnamefont {K.}~\bibnamefont {Kieu}},\ }\href {\doibase 10.1103/PhysRevA.99.063826} {\bibfield  {journal} {\bibinfo  {journal} {Physical Review A}\ }\textbf {\bibinfo {volume} {99}},\ \bibinfo {pages} {063826} (\bibinfo {year} {2019})},\ \bibinfo {note} {publisher: American Physical Society}\BibitemShut {NoStop}%
\bibitem [{\citenamefont {Kang}\ \emph {et~al.}(2009)\citenamefont {Kang}, \citenamefont {Nazarkin}, \citenamefont {Brenn},\ and\ \citenamefont {Russell}}]{kang_tightly_2009}%
  \BibitemOpen
  \bibfield  {author} {\bibinfo {author} {\bibfnamefont {M.~S.}\ \bibnamefont {Kang}}, \bibinfo {author} {\bibfnamefont {A.}~\bibnamefont {Nazarkin}}, \bibinfo {author} {\bibfnamefont {A.}~\bibnamefont {Brenn}}, \ and\ \bibinfo {author} {\bibfnamefont {P.~S.~J.}\ \bibnamefont {Russell}},\ }\href {\doibase 10.1038/nphys1217} {\bibfield  {journal} {\bibinfo  {journal} {Nature Physics}\ }\textbf {\bibinfo {volume} {5}},\ \bibinfo {pages} {276} (\bibinfo {year} {2009})},\ \bibinfo {note} {number: 4 Publisher: Nature Publishing Group}\BibitemShut {NoStop}%
\bibitem [{\citenamefont {Kang}\ \emph {et~al.}(2010)\citenamefont {Kang}, \citenamefont {Brenn},\ and\ \citenamefont {St.J.~Russell}}]{kang_all-optical_2010}%
  \BibitemOpen
  \bibfield  {author} {\bibinfo {author} {\bibfnamefont {M.~S.}\ \bibnamefont {Kang}}, \bibinfo {author} {\bibfnamefont {A.}~\bibnamefont {Brenn}}, \ and\ \bibinfo {author} {\bibfnamefont {P.}~\bibnamefont {St.J.~Russell}},\ }\href {\doibase 10.1103/PhysRevLett.105.153901} {\bibfield  {journal} {\bibinfo  {journal} {Physical Review Letters}\ }\textbf {\bibinfo {volume} {105}},\ \bibinfo {pages} {153901} (\bibinfo {year} {2010})}\BibitemShut {NoStop}%
\bibitem [{\citenamefont {Renninger}\ \emph {et~al.}(2016)\citenamefont {Renninger}, \citenamefont {Shin}, \citenamefont {Behunin}, \citenamefont {Kharel}, \citenamefont {Kittlaus},\ and\ \citenamefont {Rakich}}]{renninger_forward_2016}%
  \BibitemOpen
  \bibfield  {author} {\bibinfo {author} {\bibfnamefont {W.~H.}\ \bibnamefont {Renninger}}, \bibinfo {author} {\bibfnamefont {H.}~\bibnamefont {Shin}}, \bibinfo {author} {\bibfnamefont {R.~O.}\ \bibnamefont {Behunin}}, \bibinfo {author} {\bibfnamefont {P.}~\bibnamefont {Kharel}}, \bibinfo {author} {\bibfnamefont {E.~A.}\ \bibnamefont {Kittlaus}}, \ and\ \bibinfo {author} {\bibfnamefont {P.~T.}\ \bibnamefont {Rakich}},\ }\href {\doibase 10.1088/1367-2630/18/2/025008} {\bibfield  {journal} {\bibinfo  {journal} {New Journal of Physics}\ }\textbf {\bibinfo {volume} {18}},\ \bibinfo {pages} {025008} (\bibinfo {year} {2016})},\ \bibinfo {note} {publisher: IOP Publishing}\BibitemShut {NoStop}%
\bibitem [{\citenamefont {Zhang}\ \emph {et~al.}(2017)\citenamefont {Zhang}, \citenamefont {Wang}, \citenamefont {Cheng},\ and\ \citenamefont {Tsang}}]{zhang_forward_2017}%
  \BibitemOpen
  \bibfield  {author} {\bibinfo {author} {\bibfnamefont {Y.}~\bibnamefont {Zhang}}, \bibinfo {author} {\bibfnamefont {L.}~\bibnamefont {Wang}}, \bibinfo {author} {\bibfnamefont {Z.}~\bibnamefont {Cheng}}, \ and\ \bibinfo {author} {\bibfnamefont {H.~K.}\ \bibnamefont {Tsang}},\ }\href {\doibase 10.1063/1.4996367} {\bibfield  {journal} {\bibinfo  {journal} {Applied Physics Letters}\ }\textbf {\bibinfo {volume} {111}},\ \bibinfo {pages} {041104} (\bibinfo {year} {2017})}\BibitemShut {NoStop}%
\bibitem [{\citenamefont {Bahl}\ \emph {et~al.}(2013)\citenamefont {Bahl}, \citenamefont {Kim}, \citenamefont {Lee}, \citenamefont {Liu}, \citenamefont {Fan},\ and\ \citenamefont {Carmon}}]{bahl_brillouin_2013}%
  \BibitemOpen
  \bibfield  {author} {\bibinfo {author} {\bibfnamefont {G.}~\bibnamefont {Bahl}}, \bibinfo {author} {\bibfnamefont {K.~H.}\ \bibnamefont {Kim}}, \bibinfo {author} {\bibfnamefont {W.}~\bibnamefont {Lee}}, \bibinfo {author} {\bibfnamefont {J.}~\bibnamefont {Liu}}, \bibinfo {author} {\bibfnamefont {X.}~\bibnamefont {Fan}}, \ and\ \bibinfo {author} {\bibfnamefont {T.}~\bibnamefont {Carmon}},\ }\href {\doibase 10.1038/ncomms2994} {\bibfield  {journal} {\bibinfo  {journal} {Nature Communications}\ }\textbf {\bibinfo {volume} {4}},\ \bibinfo {pages} {1994} (\bibinfo {year} {2013})},\ \bibinfo {note} {publisher: Nature Publishing Group}\BibitemShut {NoStop}%
\bibitem [{\citenamefont {Shin}\ \emph {et~al.}(2015)\citenamefont {Shin}, \citenamefont {Cox}, \citenamefont {Jarecki}, \citenamefont {Starbuck}, \citenamefont {Wang},\ and\ \citenamefont {Rakich}}]{shin_control_2015}%
  \BibitemOpen
  \bibfield  {author} {\bibinfo {author} {\bibfnamefont {H.}~\bibnamefont {Shin}}, \bibinfo {author} {\bibfnamefont {J.~A.}\ \bibnamefont {Cox}}, \bibinfo {author} {\bibfnamefont {R.}~\bibnamefont {Jarecki}}, \bibinfo {author} {\bibfnamefont {A.}~\bibnamefont {Starbuck}}, \bibinfo {author} {\bibfnamefont {Z.}~\bibnamefont {Wang}}, \ and\ \bibinfo {author} {\bibfnamefont {P.~T.}\ \bibnamefont {Rakich}},\ }\href {\doibase 10.1038/ncomms7427} {\bibfield  {journal} {\bibinfo  {journal} {Nature Communications}\ }\textbf {\bibinfo {volume} {6}},\ \bibinfo {pages} {6427} (\bibinfo {year} {2015})},\ \bibinfo {note} {publisher: Nature Publishing Group}\BibitemShut {NoStop}%
\bibitem [{\citenamefont {Shin}\ \emph {et~al.}(2013)\citenamefont {Shin}, \citenamefont {Qiu}, \citenamefont {Jarecki}, \citenamefont {Cox}, \citenamefont {Olsson}, \citenamefont {Starbuck}, \citenamefont {Wang},\ and\ \citenamefont {Rakich}}]{shin_tailorable_2013}%
  \BibitemOpen
  \bibfield  {author} {\bibinfo {author} {\bibfnamefont {H.}~\bibnamefont {Shin}}, \bibinfo {author} {\bibfnamefont {W.}~\bibnamefont {Qiu}}, \bibinfo {author} {\bibfnamefont {R.}~\bibnamefont {Jarecki}}, \bibinfo {author} {\bibfnamefont {J.~A.}\ \bibnamefont {Cox}}, \bibinfo {author} {\bibfnamefont {R.~H.}\ \bibnamefont {Olsson}}, \bibinfo {author} {\bibfnamefont {A.}~\bibnamefont {Starbuck}}, \bibinfo {author} {\bibfnamefont {Z.}~\bibnamefont {Wang}}, \ and\ \bibinfo {author} {\bibfnamefont {P.~T.}\ \bibnamefont {Rakich}},\ }\href {\doibase 10.1038/ncomms2943} {\bibfield  {journal} {\bibinfo  {journal} {Nature Communications}\ }\textbf {\bibinfo {volume} {4}},\ \bibinfo {pages} {1944} (\bibinfo {year} {2013})},\ \bibinfo {note} {number: 1 Publisher: Nature Publishing Group}\BibitemShut {NoStop}%
\bibitem [{\citenamefont {Kittlaus}\ \emph {et~al.}(2016)\citenamefont {Kittlaus}, \citenamefont {Shin},\ and\ \citenamefont {Rakich}}]{kittlaus_large_2016}%
  \BibitemOpen
  \bibfield  {author} {\bibinfo {author} {\bibfnamefont {E.~A.}\ \bibnamefont {Kittlaus}}, \bibinfo {author} {\bibfnamefont {H.}~\bibnamefont {Shin}}, \ and\ \bibinfo {author} {\bibfnamefont {P.~T.}\ \bibnamefont {Rakich}},\ }\href {\doibase 10.1038/nphoton.2016.112} {\bibfield  {journal} {\bibinfo  {journal} {Nature Photonics}\ }\textbf {\bibinfo {volume} {10}},\ \bibinfo {pages} {463} (\bibinfo {year} {2016})},\ \bibinfo {note} {publisher: Nature Publishing Group}\BibitemShut {NoStop}%
\bibitem [{\citenamefont {Bahl}\ \emph {et~al.}(2011)\citenamefont {Bahl}, \citenamefont {Zehnpfennig}, \citenamefont {Tomes},\ and\ \citenamefont {Carmon}}]{bahl_stimulated_2011}%
  \BibitemOpen
  \bibfield  {author} {\bibinfo {author} {\bibfnamefont {G.}~\bibnamefont {Bahl}}, \bibinfo {author} {\bibfnamefont {J.}~\bibnamefont {Zehnpfennig}}, \bibinfo {author} {\bibfnamefont {M.}~\bibnamefont {Tomes}}, \ and\ \bibinfo {author} {\bibfnamefont {T.}~\bibnamefont {Carmon}},\ }\href {\doibase 10.1038/ncomms1412} {\bibfield  {journal} {\bibinfo  {journal} {Nature Communications}\ }\textbf {\bibinfo {volume} {2}},\ \bibinfo {pages} {403} (\bibinfo {year} {2011})},\ \bibinfo {note} {publisher: Nature Publishing Group}\BibitemShut {NoStop}%
\bibitem [{\citenamefont {Yu}\ \emph {et~al.}(2022)\citenamefont {Yu}, \citenamefont {Shen}, \citenamefont {Yang}, \citenamefont {Qi}, \citenamefont {Jiang}, \citenamefont {Brambilla}, \citenamefont {Dong},\ and\ \citenamefont {Wang}}]{yu_investigation_2022}%
  \BibitemOpen
  \bibfield  {author} {\bibinfo {author} {\bibfnamefont {J.}~\bibnamefont {Yu}}, \bibinfo {author} {\bibfnamefont {Z.}~\bibnamefont {Shen}}, \bibinfo {author} {\bibfnamefont {Z.}~\bibnamefont {Yang}}, \bibinfo {author} {\bibfnamefont {S.}~\bibnamefont {Qi}}, \bibinfo {author} {\bibfnamefont {Y.}~\bibnamefont {Jiang}}, \bibinfo {author} {\bibfnamefont {G.}~\bibnamefont {Brambilla}}, \bibinfo {author} {\bibfnamefont {C.-H.}\ \bibnamefont {Dong}}, \ and\ \bibinfo {author} {\bibfnamefont {P.}~\bibnamefont {Wang}},\ }\href {\doibase 10.1109/JPHOT.2022.3145033} {\bibfield  {journal} {\bibinfo  {journal} {IEEE Photonics Journal}\ }\textbf {\bibinfo {volume} {14}},\ \bibinfo {pages} {1} (\bibinfo {year} {2022})},\ \bibinfo {note} {conference Name: IEEE Photonics Journal}\BibitemShut {NoStop}%
\bibitem [{\citenamefont {Wang}\ \emph {et~al.}(2024)\citenamefont {Wang}, \citenamefont {Hu}, \citenamefont {Lao}, \citenamefont {Wang}, \citenamefont {Jin}, \citenamefont {Zhou}, \citenamefont {Lei}, \citenamefont {Wang}, \citenamefont {Liu}, \citenamefont {Yang},\ and\ \citenamefont {Li}}]{wang_taming_2024}%
  \BibitemOpen
  \bibfield  {author} {\bibinfo {author} {\bibfnamefont {M.}~\bibnamefont {Wang}}, \bibinfo {author} {\bibfnamefont {Z.-G.}\ \bibnamefont {Hu}}, \bibinfo {author} {\bibfnamefont {C.}~\bibnamefont {Lao}}, \bibinfo {author} {\bibfnamefont {Y.}~\bibnamefont {Wang}}, \bibinfo {author} {\bibfnamefont {X.}~\bibnamefont {Jin}}, \bibinfo {author} {\bibfnamefont {X.}~\bibnamefont {Zhou}}, \bibinfo {author} {\bibfnamefont {Y.}~\bibnamefont {Lei}}, \bibinfo {author} {\bibfnamefont {Z.}~\bibnamefont {Wang}}, \bibinfo {author} {\bibfnamefont {W.}~\bibnamefont {Liu}}, \bibinfo {author} {\bibfnamefont {Q.-F.}\ \bibnamefont {Yang}}, \ and\ \bibinfo {author} {\bibfnamefont {B.-B.}\ \bibnamefont {Li}},\ }\href {\doibase 10.1103/PhysRevX.14.011056} {\bibfield  {journal} {\bibinfo  {journal} {Physical Review X}\ }\textbf {\bibinfo {volume} {14}},\ \bibinfo {pages} {011056} (\bibinfo {year} {2024})},\ \bibinfo {note} {publisher: American Physical Society}\BibitemShut {NoStop}%
\bibitem [{\citenamefont {Lee}\ and\ \citenamefont {Agrawal}(2003)}]{lee_suppression_2003}%
  \BibitemOpen
  \bibfield  {author} {\bibinfo {author} {\bibfnamefont {H.}~\bibnamefont {Lee}}\ and\ \bibinfo {author} {\bibfnamefont {G.~P.}\ \bibnamefont {Agrawal}},\ }\href {\doibase 10.1364/OE.11.003467} {\bibfield  {journal} {\bibinfo  {journal} {Optics Express}\ }\textbf {\bibinfo {volume} {11}},\ \bibinfo {pages} {3467} (\bibinfo {year} {2003})},\ \bibinfo {note} {publisher: Optica Publishing Group}\BibitemShut {NoStop}%
\bibitem [{\citenamefont {Merklein}\ \emph {et~al.}(2015)\citenamefont {Merklein}, \citenamefont {Kabakova}, \citenamefont {Büttner}, \citenamefont {Choi}, \citenamefont {Luther-Davies}, \citenamefont {Madden},\ and\ \citenamefont {Eggleton}}]{merklein_enhancing_2015}%
  \BibitemOpen
  \bibfield  {author} {\bibinfo {author} {\bibfnamefont {M.}~\bibnamefont {Merklein}}, \bibinfo {author} {\bibfnamefont {I.~V.}\ \bibnamefont {Kabakova}}, \bibinfo {author} {\bibfnamefont {T.~F.~S.}\ \bibnamefont {Büttner}}, \bibinfo {author} {\bibfnamefont {D.-Y.}\ \bibnamefont {Choi}}, \bibinfo {author} {\bibfnamefont {B.}~\bibnamefont {Luther-Davies}}, \bibinfo {author} {\bibfnamefont {S.~J.}\ \bibnamefont {Madden}}, \ and\ \bibinfo {author} {\bibfnamefont {B.~J.}\ \bibnamefont {Eggleton}},\ }\href {\doibase 10.1038/ncomms7396} {\bibfield  {journal} {\bibinfo  {journal} {Nature Communications}\ }\textbf {\bibinfo {volume} {6}},\ \bibinfo {pages} {6396} (\bibinfo {year} {2015})},\ \bibinfo {note} {publisher: Nature Publishing Group}\BibitemShut {NoStop}%
\bibitem [{\citenamefont {Ibsen}\ \emph {et~al.}(1998)\citenamefont {Ibsen}, \citenamefont {Durkin}, \citenamefont {Cole},\ and\ \citenamefont {Laming}}]{ibsen_sinc-sampled_1998}%
  \BibitemOpen
  \bibfield  {author} {\bibinfo {author} {\bibfnamefont {M.}~\bibnamefont {Ibsen}}, \bibinfo {author} {\bibfnamefont {M.}~\bibnamefont {Durkin}}, \bibinfo {author} {\bibfnamefont {M.}~\bibnamefont {Cole}}, \ and\ \bibinfo {author} {\bibfnamefont {R.}~\bibnamefont {Laming}},\ }\href {\doibase 10.1109/68.681504} {\bibfield  {journal} {\bibinfo  {journal} {IEEE Photonics Technology Letters}\ }\textbf {\bibinfo {volume} {10}},\ \bibinfo {pages} {842} (\bibinfo {year} {1998})},\ \bibinfo {note} {conference Name: IEEE Photonics Technology Letters}\BibitemShut {NoStop}%
\bibitem [{\citenamefont {{K. J. Satzinger}}\ \emph {et~al.}(2018)\citenamefont {{K. J. Satzinger}}, \citenamefont {Zhong}, \citenamefont {Chang}, \citenamefont {Peairs}, \citenamefont {Bienfait}, \citenamefont {Chou}, \citenamefont {Cleland}, \citenamefont {Conner}, \citenamefont {Dumur}, \citenamefont {Grebel}, \citenamefont {Gutierrez}, \citenamefont {November}, \citenamefont {Povey}, \citenamefont {Whiteley}, \citenamefont {Awschalom}, \citenamefont {Schuster},\ and\ \citenamefont {Cleland}}]{k_j_satzinger_quantum_2018}%
  \BibitemOpen
  \bibfield  {author} {\bibinfo {author} {\bibnamefont {{K. J. Satzinger}}}, \bibinfo {author} {\bibfnamefont {Y.~P.}\ \bibnamefont {Zhong}}, \bibinfo {author} {\bibfnamefont {H.-S.}\ \bibnamefont {Chang}}, \bibinfo {author} {\bibfnamefont {G.~A.}\ \bibnamefont {Peairs}}, \bibinfo {author} {\bibfnamefont {A.}~\bibnamefont {Bienfait}}, \bibinfo {author} {\bibfnamefont {M.-H.}\ \bibnamefont {Chou}}, \bibinfo {author} {\bibfnamefont {A.~Y.}\ \bibnamefont {Cleland}}, \bibinfo {author} {\bibfnamefont {C.~R.}\ \bibnamefont {Conner}}, \bibinfo {author} {\bibfnamefont {Ã.}~\bibnamefont {Dumur}}, \bibinfo {author} {\bibfnamefont {J.}~\bibnamefont {Grebel}}, \bibinfo {author} {\bibfnamefont {I.}~\bibnamefont {Gutierrez}}, \bibinfo {author} {\bibfnamefont {B.~H.}\ \bibnamefont {November}}, \bibinfo {author} {\bibfnamefont {R.~G.}\ \bibnamefont {Povey}}, \bibinfo {author} {\bibfnamefont {S.~J.}\ \bibnamefont {Whiteley}}, \bibinfo {author} {\bibfnamefont {D.~D.}\ \bibnamefont {Awschalom}}, \bibinfo {author} {\bibfnamefont
  {D.~I.}\ \bibnamefont {Schuster}}, \ and\ \bibinfo {author} {\bibfnamefont {A.~N.}\ \bibnamefont {Cleland}},\ }\href {\doibase 10.1038/s41586-018-0719-5} {\bibfield  {journal} {\bibinfo  {journal} {Nature}\ }\textbf {\bibinfo {volume} {563}},\ \bibinfo {pages} {661} (\bibinfo {year} {2018})},\ \bibinfo {note} {number: 7733 Publisher: Nature Publishing Group}\BibitemShut {NoStop}%
\bibitem [{\citenamefont {Galland}\ \emph {et~al.}(2014)\citenamefont {Galland}, \citenamefont {Sangouard}, \citenamefont {Piro}, \citenamefont {Gisin},\ and\ \citenamefont {Kippenberg}}]{galland_heralded_2014}%
  \BibitemOpen
  \bibfield  {author} {\bibinfo {author} {\bibfnamefont {C.}~\bibnamefont {Galland}}, \bibinfo {author} {\bibfnamefont {N.}~\bibnamefont {Sangouard}}, \bibinfo {author} {\bibfnamefont {N.}~\bibnamefont {Piro}}, \bibinfo {author} {\bibfnamefont {N.}~\bibnamefont {Gisin}}, \ and\ \bibinfo {author} {\bibfnamefont {T.~J.}\ \bibnamefont {Kippenberg}},\ }\href {\doibase 10.1103/PhysRevLett.112.143602} {\bibfield  {journal} {\bibinfo  {journal} {Physical Review Letters}\ }\textbf {\bibinfo {volume} {112}},\ \bibinfo {pages} {143602} (\bibinfo {year} {2014})}\BibitemShut {NoStop}%
\bibitem [{\citenamefont {Hong}\ \emph {et~al.}(2017)\citenamefont {Hong}, \citenamefont {Riedinger}, \citenamefont {Marinković}, \citenamefont {Wallucks}, \citenamefont {Hofer}, \citenamefont {Norte}, \citenamefont {Aspelmeyer},\ and\ \citenamefont {Gröblacher}}]{hong_hanbury_2017}%
  \BibitemOpen
  \bibfield  {author} {\bibinfo {author} {\bibfnamefont {S.}~\bibnamefont {Hong}}, \bibinfo {author} {\bibfnamefont {R.}~\bibnamefont {Riedinger}}, \bibinfo {author} {\bibfnamefont {I.}~\bibnamefont {Marinković}}, \bibinfo {author} {\bibfnamefont {A.}~\bibnamefont {Wallucks}}, \bibinfo {author} {\bibfnamefont {S.~G.}\ \bibnamefont {Hofer}}, \bibinfo {author} {\bibfnamefont {R.~A.}\ \bibnamefont {Norte}}, \bibinfo {author} {\bibfnamefont {M.}~\bibnamefont {Aspelmeyer}}, \ and\ \bibinfo {author} {\bibfnamefont {S.}~\bibnamefont {Gröblacher}},\ }\href {\doibase 10.1126/science.aan7939} {\bibfield  {journal} {\bibinfo  {journal} {Science}\ }\textbf {\bibinfo {volume} {358}},\ \bibinfo {pages} {203} (\bibinfo {year} {2017})}\BibitemShut {NoStop}%
\bibitem [{\citenamefont {Rakhubovsky}\ and\ \citenamefont {Filip}(2017)}]{rakhubovsky_photon-phonon-photon_2017}%
  \BibitemOpen
  \bibfield  {author} {\bibinfo {author} {\bibfnamefont {A.~A.}\ \bibnamefont {Rakhubovsky}}\ and\ \bibinfo {author} {\bibfnamefont {R.}~\bibnamefont {Filip}},\ }\href {\doibase 10.1038/srep46764} {\bibfield  {journal} {\bibinfo  {journal} {Scientific Reports}\ }\textbf {\bibinfo {volume} {7}},\ \bibinfo {pages} {46764} (\bibinfo {year} {2017})},\ \bibinfo {note} {publisher: Nature Publishing Group}\BibitemShut {NoStop}%
\bibitem [{\citenamefont {Aspelmeyer}\ \emph {et~al.}(2014)\citenamefont {Aspelmeyer}, \citenamefont {Kippenberg},\ and\ \citenamefont {Marquardt}}]{aspelmeyer_cavity_2014}%
  \BibitemOpen
  \bibfield  {author} {\bibinfo {author} {\bibfnamefont {M.}~\bibnamefont {Aspelmeyer}}, \bibinfo {author} {\bibfnamefont {T.~J.}\ \bibnamefont {Kippenberg}}, \ and\ \bibinfo {author} {\bibfnamefont {F.}~\bibnamefont {Marquardt}},\ }\href {\doibase 10.1103/RevModPhys.86.1391} {\bibfield  {journal} {\bibinfo  {journal} {Reviews of Modern Physics}\ }\textbf {\bibinfo {volume} {86}},\ \bibinfo {pages} {1391} (\bibinfo {year} {2014})},\ \bibinfo {note} {publisher: American Physical Society}\BibitemShut {NoStop}%
\bibitem [{\citenamefont {Doeleman}\ \emph {et~al.}(2023)\citenamefont {Doeleman}, \citenamefont {Schatteburg}, \citenamefont {Benevides}, \citenamefont {Vollenweider}, \citenamefont {Macri},\ and\ \citenamefont {Chu}}]{doeleman_brillouin_2023}%
  \BibitemOpen
  \bibfield  {author} {\bibinfo {author} {\bibfnamefont {H.~M.}\ \bibnamefont {Doeleman}}, \bibinfo {author} {\bibfnamefont {T.}~\bibnamefont {Schatteburg}}, \bibinfo {author} {\bibfnamefont {R.}~\bibnamefont {Benevides}}, \bibinfo {author} {\bibfnamefont {S.}~\bibnamefont {Vollenweider}}, \bibinfo {author} {\bibfnamefont {D.}~\bibnamefont {Macri}}, \ and\ \bibinfo {author} {\bibfnamefont {Y.}~\bibnamefont {Chu}},\ }\href {\doibase 10.1103/PhysRevResearch.5.043140} {\bibfield  {journal} {\bibinfo  {journal} {Physical Review Research}\ }\textbf {\bibinfo {volume} {5}},\ \bibinfo {pages} {043140} (\bibinfo {year} {2023})}\BibitemShut {NoStop}%
\bibitem [{\citenamefont {Liu}\ \emph {et~al.}(2022)\citenamefont {Liu}, \citenamefont {Jin}, \citenamefont {Cheng}, \citenamefont {Chauhan}, \citenamefont {Puckett}, \citenamefont {Nelson}, \citenamefont {Behunin}, \citenamefont {Rakich},\ and\ \citenamefont {Blumenthal}}]{liu_ultralow_2022}%
  \BibitemOpen
  \bibfield  {author} {\bibinfo {author} {\bibfnamefont {K.}~\bibnamefont {Liu}}, \bibinfo {author} {\bibfnamefont {N.}~\bibnamefont {Jin}}, \bibinfo {author} {\bibfnamefont {H.}~\bibnamefont {Cheng}}, \bibinfo {author} {\bibfnamefont {N.}~\bibnamefont {Chauhan}}, \bibinfo {author} {\bibfnamefont {M.~W.}\ \bibnamefont {Puckett}}, \bibinfo {author} {\bibfnamefont {K.~D.}\ \bibnamefont {Nelson}}, \bibinfo {author} {\bibfnamefont {R.~O.}\ \bibnamefont {Behunin}}, \bibinfo {author} {\bibfnamefont {P.~T.}\ \bibnamefont {Rakich}}, \ and\ \bibinfo {author} {\bibfnamefont {D.~J.}\ \bibnamefont {Blumenthal}},\ }\href {\doibase 10.1364/OL.454392} {\bibfield  {journal} {\bibinfo  {journal} {Optics Letters}\ }\textbf {\bibinfo {volume} {47}},\ \bibinfo {pages} {1855} (\bibinfo {year} {2022})},\ \bibinfo {note} {publisher: Optica Publishing Group}\BibitemShut {NoStop}%
\bibitem [{\citenamefont {Ma}\ \emph {et~al.}(2020)\citenamefont {Ma}, \citenamefont {Chen}, \citenamefont {Li}, \citenamefont {Tang}, \citenamefont {Sua}, \citenamefont {Fan},\ and\ \citenamefont {Huang}}]{ma_ultrabright_2020}%
  \BibitemOpen
  \bibfield  {author} {\bibinfo {author} {\bibfnamefont {Z.}~\bibnamefont {Ma}}, \bibinfo {author} {\bibfnamefont {J.-Y.}\ \bibnamefont {Chen}}, \bibinfo {author} {\bibfnamefont {Z.}~\bibnamefont {Li}}, \bibinfo {author} {\bibfnamefont {C.}~\bibnamefont {Tang}}, \bibinfo {author} {\bibfnamefont {Y.~M.}\ \bibnamefont {Sua}}, \bibinfo {author} {\bibfnamefont {H.}~\bibnamefont {Fan}}, \ and\ \bibinfo {author} {\bibfnamefont {Y.-P.}\ \bibnamefont {Huang}},\ }\href {\doibase 10.1103/PhysRevLett.125.263602} {\bibfield  {journal} {\bibinfo  {journal} {Physical Review Letters}\ }\textbf {\bibinfo {volume} {125}},\ \bibinfo {pages} {263602} (\bibinfo {year} {2020})}\BibitemShut {NoStop}%
\end{thebibliography}
\end{document}